\RequirePackage{fix-cm}
\documentclass[twocolumn, epjc3, comma, sort&compress, natbib]{svjour3}
\bibpunct{[}{]}{,}{n}{}{,} 

\journalname{Eur. Phys. J. A}

\usepackage{latexsym}
\usepackage{amsmath}
\usepackage{amssymb}
\usepackage{amsfonts}
\usepackage{CJKutf8}
\usepackage{enumitem}

\usepackage[mathscr,scaled=1.15]{urwchancal}
\DeclareFontFamily{OT1}{pzc}{}
\DeclareFontShape{OT1}{pzc}{m}{it}%
{<-> s * [1.15] pzcmi7t}{}
\DeclareMathAlphabet{\mathpzc}{OT1}{pzc}{m}{it}

\usepackage{color}

\usepackage{supertabular}
\usepackage{placeins}
\usepackage{epsfig}
\usepackage{graphicx}

\definecolor{purple}{rgb}{0.5,0,0.5}
\definecolor{blue}{rgb}{0.0,0,0.9}
\definecolor{prdblue}{rgb}{0.133,0.118,0.498}
\usepackage[colorlinks=true, pdfstartview=FitV, linkcolor=prdblue, citecolor= prdblue, urlcolor=prdblue]{hyperref}

\RequirePackage{comment}

\begin{document}
\begin{CJK*}{UTF8}{gbsn}

\title{$\,$\\[-6ex]\hspace*{\fill}{\normalsize{\sf\emph{Preprint no}.\
NJU-INP 110-25}}\\[1ex]
Symmetry-preserving calculation \\ of pion light-front wave functions}

\author{
Z.-Q.\ Yao (姚照千)\thanksref{HZDR}%
       $^{,\href{https://orcid.org/0000-0002-9621-6994}{\textcolor[rgb]{0.00,1.00,0.00}{\sf ID}}}$
\and
Z.-N.\ Xu (徐珍妮)\thanksref{UHe}%
    $^{,\href{https://orcid.org/0000-0002-9104-9680}{\textcolor[rgb]{0.00,1,0.00}{\sf ID}}}$
\and
Y.-Y.\ Xiao (肖宇洋)\thanksref{NJU,INP}%
       $^{,\href{https://orcid.org/0009-0006-1963-7388}{\textcolor[rgb]{0.00,1.00,0.00}{\sf ID}}}$
\and     \\C.\ D.\ Roberts\thanksref{NJU,INP}%
       $^{,\href{https://orcid.org/0000-0002-2937-1361}{\textcolor[rgb]{0.00,1.00,0.00}{\sf ID}}}$
\and
J.\ Rodr\'iguez-Quintero\thanksref{UHe}%
       $^{,\href{https://orcid.org/0000-0002-1651-5717}{\textcolor[rgb]{0.00,1,0.00}{\sf ID}}}$}

\authorrunning{Z.-Q.\ Yao \emph{et al}.} 

\institute{
\href{https://ror.org/01zy2cs03}{Helmholtz-Zentrum Dresden-Rossendorf}, Bautzner Landstra{\ss}e 400, D-01328 Dresden, Germany \label{HZDR}
\and
Department of Integrated Sciences and Center for Advanced Studies in Physics, Mathematics and Computation, \href{https://ror.org/03a1kt624}{University of Huelva}, E-21071 Huelva, Spain \label{UHe}
\and
School of Physics, \href{https://ror.org/01rxvg760}{Nanjing University}, Nanjing, Jiangsu 210093, China \label{NJU}
\and
Institute for Nonperturbative Physics, \href{https://ror.org/01rxvg760}{Nanjing University}, Nanjing, Jiangsu 210093, China \label{INP}
\\[1ex]
Email:
\href{mailto:z.yao@hzdr.de}{z.yao@hzdr.de} (ZQY);
\href{mailto:zhenni.xu@dci.uhu.es}{zhenni.xu@dci.uhu.es} (ZNX);
\href{mailto:cdroberts@nju.edu.cn}{cdroberts@nju.edu.cn} (CDR)
            }

\date{2026 February 22}  

\maketitle

\end{CJK*}

\begin{abstract}
Poincar\'e-covariant Bethe-Salpeter wave functions are used to calculate light-front wave functions (LFWFs) of the pion, $\pi$, and an analogue state, $\pi_{s\bar s}$.
The current masses of the degenerate valence constituents in the $\pi_{s\bar s}$ are around $25$-times larger than those of the pion's valence constituents.
Both valence spin-anti\-aligned ($\mathpzc L=0$) and spin-aligned ($\mathpzc L=1$) components are obtained and combined to produce the complete LFWF for each system.
Comparing predictions delivered by two distinct Bethe-Salpeter kernels, the impact of nonperturbative dynamical effects contained in the more sophisticated (bRL) kernel are seen to be significant; and contrasts between $\pi$, $\pi_{s \bar s}$ results reveal the interplay between emergent hadron mass and mass effects owing to Higgs-boson couplings.
Amongst the results, one finds that for $\pi$, $\pi_{s\bar s}$, the LFWFs can be approximated by a separable form, with that representation being pointwise reliable in the bRL cases.
Moreover, the $\mathpzc L=1$ component is important; so a LFWF obtained after omission of this piece is typically a poor representation of the  system.
These features are naturally expressed in $\pi$, $\pi_{s\bar s}$ transverse momentum dependent parton distribution functions (TMDs).
In this connection, it is found that a Gaussian \textit{Ansatz} can only provide a rough guide to TMD pointwise behaviour: magnitude deviations between \textit{Ansatz} and prediction exceed a factor of two on $k_\perp^2 \gtrsim 0.55\,$GeV$^2$.
One should therefore be cautious in interpreting conclusions drawn from phenomenological analyses based upon Gaussian \textit{Ans\"atze}.
\end{abstract}

\section{Introduction}
The physical pion is Nature's most fundamental (near) Nambu-Goldstone boson -- only the small light quark current mass disturbs its Nambu-Goldstone character \cite{Horn:2016rip}.  Owing to that small mass, it is also the lightest strongly interacting particle, \textit{viz}.\ a meson built from light valence quark + antiquark degrees of freedom.  Resolving this dichotomy has long been a goal of high-energy nuclear and particle physics.  Now, with the dawning of an era of high-energy, high-luminosity facilities \cite{Denisov:2018unj, Aguilar:2019teb, Brodsky:2020vco, Chen:2020ijn, Anderle:2021wcy, Arrington:2021biu, Quintans:2022utc, Lu:2025bnm}, it becomes realistic to think of producing detailed maps of pion structure and therewith validating modern theoretical pictures of the pion \cite{Horn:2016rip, Roberts:2021nhw}; so, an explanation of a basic feature of Nature may finally be within reach.

A path to an understanding of the structure of a quantum system is provided by the object's wave function.
The Schr\"odinger wave function associated with a nonrelativistic system is a probability amplitude.  It can be computed using a Lippmann-Schwinger equation.
In Poincar\'e-invariant quantum field theory, the equivalent of the latter is the Bethe-Salpeter equation \cite{Salpeter:1951sz}.  However, owing to the special features of such theories -- in particular, the loss of particle number conservation -- the Poincar\'e-covariant Bethe-Salpeter wave function does not admit interpretation as a probability amplitude.

In quantum field theory, a clean analogue of the Schr\"odinger wave function is provided be the light-front wave function (LFWF) \cite[Sec.\,3A]{Roberts:2021nhw}, \cite{Brodsky:1979gy}.  Notably, for any system described by a Poincar\'e-covariant wave function, ${\mathpzc X}$, the associated LFWF can be obtained via projection of ${\mathpzc X}$ onto the light front \cite{tHooft:1974pnl}.  
This is a precise statement: knowing the QCD light-front Hamiltonian is not a prerequisite for obtaining a LFWF.
In hadron physics applications, the projection approach was shown to be practicable in Ref.\,\cite{Chang:2013pq} and it has since been widely used. Herein, we adapt the scheme to the calculation of the LFWF of the pion, $\pi$, and a (fictitious) analogue state, $\pi_{s \bar s}$, built from valence degrees of freedom with degenerate current-masses that are inflated to match that of the strange, $s$, quark.
Using these LFWFs, we then proceed to deliver results for the helicity-independent transverse momentum dependent parton distribution function (TMD) of each system.
Whilst a system's LFWF is not directly accessible, it is anticipated that existing and future accelerators will deliver precise data that enable empirical inference of the pion TMD \cite{Quintans:2022utc, Arrington:2021biu}.

Over the past forty-five years, many phenomenological models of pion LFWFs have been developed and employed; recent examples are discussed, \textit{e.g}., in Refs.\,\cite{Choi:2014ifm, Xu:2018eii, dePaula:2020qna, Moita:2022lfu, Albino:2022gzs, Li:2022mlg, Pasquini:2023aaf, Zhu:2023lst, Wang:2025usl}.
In contrast, herein we employ continuum Schwinger function methods (CSMs) \cite{Roberts:2012sv, Eichmann:2016yit, Qin:2020rad} to deliver parameter-free predictions for $\pi$, $\pi_{s\bar s}$ Poincar\'e-covariant wave functions, which we subsequently project onto the light front.  A merit of this approach is the straightforward realisation of the pion as simultaneously both a Nambu-Goldstone mode and quark + antiquark bound state \cite{Maris:1997hd, Maris:1997tm, Chang:2009zb, Fischer:2009jm, Qin:2014vya}.

When employing CSMs, the wave function obtained depends on the kernel used to complete the Bethe-Salpeter equation.
We therefore compare results calculated using two different symmetry-preserving, systema\-tically-improvable kernels.
Namely, (\textsf{I}) the rainbow-lad\-der (RL) truncation, which is the leading-order approximation in the scheme introduced in Refs.\,\cite{Munczek:1994zz, Bender:1996bb}; and (\textsf{II}) the nonperturbatively constructed extension (bRL) of that truncation elucidated and employed in Refs.\,\cite{Qin:2020jig, Xu:2022kng, Xu:2025cyj}.
These comparisons enable one to \cite{Roberts:2020udq, Roberts:2021nhw, Binosi:2022djx, Ding:2022ows, Ferreira:2023fva, Raya:2024ejx, Achenbach:2025kfx}: highlight the expression of emergent hadron mass (EHM) in the pion LFWF and TMD; and consider aspects of the impact of constructive interference between EHM and Higgs-boson mass generating effects in the Standard Model.

It is worth recording that in the present context, as highlighted elsewhere \cite{Qin:2020jig, Xu:2022kng}, symmetry-preserving means that no symmetries are broken by the kernel approximation, \emph{e.g}., amongst others, all the following are preserved: the Euclidean equivalent of Lorentz invariance, translation invariance, and the Ward-Green-Takahashi identities for vector and axialvector currents.
This feature (symmetry-preservation) is of great value when calculating LFWFs because it means, for instance, that the Bethe-Salpeter wave function obtained with such a kernel can truly be equated with that of a (near) Nambu-Goldstone and that the subsequent light-front projection is rigorously well defined.

We stress that the results discussed herein are to be interpreted as expressing meson structural properties at the hadron scale, $\zeta_{\cal H}<m_N$, $m_N$ is the nucleon mass.
At $\zeta_{\cal H}$, all properties of the hadron are carried by its quasiparticle valence degrees of freedom.  Evolution to higher scales can be accomplished using the all-orders (AO) scheme, detailed in Ref.\,\cite{Yin:2023dbw}, which has proven efficacious in numerous applications.

Section~\ref{Sec2} sketches the CSM calculation of meson Bethe-Salpeter amplitudes.
Algebraic connections between these amplitudes and meson LFWFs are provided in Sec.\,\ref{Sec3}.
A practical procedure for projecting the Bethe-Salpeter amplitudes onto the light-front is explained and employed in Sec.\,\ref{Sec4}.
Section~\ref{Sec5} discusses the helicity-independent TMDs determined by the LFWFs thus obtained.
Section~\ref{Sec6} provides a summary and perspective.

\section{Poincar\'e-covariant Bethe–Salpeter wave functions}
\label{Sec2}
Using CSMs, meson bound state problems are solved by considering a set of coupled gap and Bethe-Salpeter equations \cite{Roberts:2012sv}.
Practical implementations of the RL truncation can be traced from Ref.\,\cite{Maris:1997tm} and first steps beyond RL truncation are described in Refs.\,\cite{Chang:2009zb, Fischer:2009jm}.
In either case, two elements are key, \textit{viz}.\
(\textit{a}) the effective charge;
and
(\textit{b}) the dressed gluon-quark vertex, $\Gamma_\nu$.

As explained in Refs.\,\cite{Qin:2011dd, Binosi:2014aea, Binosi:2016wcx, Yao:2024ixu}, studies of the gauge sector in quantum chromodynamics (QCD) lead one to the following practicable form for the product of effective charge and gluon $2$-point Schwinger function:
\begin{align}
\label{defcalG}
 \tilde{\mathpzc A}(y) & =
 \frac{2\pi}{\omega^4} D e^{-y/\omega^2} + \frac{2\pi \gamma_m \mathcal{F}(y)}{\ln\big[ \tau+(1+y/\Lambda_{\rm QCD}^2)^2 \big]}\,,
\end{align}
where $\gamma_m=12/25$, $\Lambda_{\rm QCD} = 0.234\,$GeV, $\tau={\rm e}^2-1$, and ${\cal F}(y) = \{1 - \exp(-y/\Lambda_{\mathpzc I}^2)\}/y$, $\Lambda_{\mathpzc I}=1\,$GeV.

Using Eq.\,\eqref{defcalG}, the kernel of the dressed-quark gap equation has the form $(l=(p-q), y=l^2)$ \cite{Maris:1997tm, Binosi:2016wcx}:
\begin{equation}
\label{DSEkernel}
\tilde{\mathpzc A}(y) T_{\mu\nu}(l) [i\gamma_\mu\frac{\lambda^{a}}{2} ]_{tr} [i\Gamma_\nu(q,p)\frac{\lambda^{a}}{2} ]_{su}\,,
\end{equation}
where $p,q$ are quark momenta before and after interaction and $l^2 T_{\mu\nu}(l) = l^2 \delta_{\mu\nu} - l_\mu l_\nu$.  This tensor structure specifies Landau gauge, employed because, \emph{inter alia}, it is a fixed point of the QCD renormalisation group.
In solving all relevant Schwinger function equations, we use a mass-independent (chiral-limit) momen\-tum-subtraction renormalisation scheme \cite{Chang:2008ec}, with re\-normalisation scale $\zeta=19\,$GeV.  At this scale, the quark wave function renormalisation constants are practically unity, which is a practical merit.  Evolution to different scales is straightforward.


\begin{table}[t]
\caption{ \label{params}
{\sf Panel A}.
Quark current masses and interaction parameters, Eq.\,\eqref{defcalG}, used to calculate meson wave functions: rainbow ladder (RL); and beyond rainbow-ladder (bRL), Eq.\,\eqref{EqbRL}.
The strength of the bRL anomalous chromomagnetic moment is fixed by ${\mathpzc a} = 1.1$.
{\sf Panel B}.
Meson masses and leptonic decay constants obtained using the values in Panel A.
Evolved at leading order to $\zeta = 2\,$GeV, the listed renormalisation group invariant current masses correspond to the following values (in GeV):
RL -- $0.0054, 0.121$; bRL -- $0.0031, 0.083$.
For context, Ref.\,\cite[PDG]{ParticleDataGroup:2024cfk} lists the following results: $0.0035$, $0.094$, values which align well with the bRL masses.
%
%
(All tabulated quantities listed in GeV.)
}
\begin{tabular*}
{\hsize}
{
l@{\extracolsep{0ptplus1fil}}
|l@{\extracolsep{0ptplus1fil}}
l@{\extracolsep{0ptplus1fil}}
l@{\extracolsep{0ptplus1fil}}
l@{\extracolsep{0ptplus1fil}}}\hline\hline
\centering
{\sf A} & $\hat m$ & $\hat m_s$ & $\omega$ & $(\omega D)^{1/3}$ \\ \hline
RL  & $0.0072$ & $0.161$ & $0.5$ & $ 0.80$\\
bRL $\ $ & $0.0041$ & $0.110$ & $0.8$ & $ 0.72$\\
\hline
\end{tabular*}

\medskip

\begin{tabular*}
{\hsize}
{
l@{\extracolsep{0ptplus1fil}}
|l@{\extracolsep{0ptplus1fil}}
l@{\extracolsep{0ptplus1fil}}
l@{\extracolsep{0ptplus1fil}}
l@{\extracolsep{0ptplus1fil}}}\hline
\centering
{\sf B} & $m_\pi$ & $m_{\pi_{s\bar s}}$ & $f_\pi$ & $f_{\pi_{s\bar s}}$ \\ \hline
RL  & $0.14$ & $0.69$ & $0.093$ & $ 0.134\phantom{00}$\\
bRL $\ $ & $0.14$ & $0.67$ & $0.098$ & $ 0.117$\\
\hline\hline
\end{tabular*}

\end{table}

In the context of Eq.\,\eqref{DSEkernel}, one may introduce RL truncation as \cite[Sec.\,II]{Maris:1997tm} $\Gamma_\nu^{\rm RL}(q,p) = \gamma_\nu$; then the EHM-improved bRL vertex employed in Refs.\,\cite{Xu:2022kng, Xu:2025cyj} takes the following form:
\begin{equation}
\label{EqbRL}
\Gamma_\nu(q,p) =
\Gamma_\nu^{\rm RL}(q,p)
- {\mathpzc a} \, \kappa(y) \, \sigma_{\alpha\nu} \, l_\alpha \,,
\end{equation}
where ${\mathpzc a} = 1.1$ measures the strength of the associated EHM-induced dressed-quark anomalous chromomagnetic moment (ACM) term \cite{Singh:1985sg, Bicudo:1998qb, Chang:2010hb, Qin:2013mta}.
Since the ACM is nonperturbative, an IR-focused profile function is appropriate, \textit{viz}.\ $\kappa(y) = (1/\omega) \exp(-y/\omega^2)$.
Notably, in the hadron spectrum, the ACM improvement eliminates most defects of RL truncation results \cite{Xu:2022kng} and it is also known to have a discernible impact on hadron structure \cite{Xu:2025cyj}.

Using the quark current masses and parameter values in Table~\ref{params}\,A, one can solve the necessary gap and Bethe-Salpeter equations using what are now standard algorithms \cite{Maris:1997tm, Maris:2005tt, Krassnigg:2009gd}.
This yields the Poincar\'e-covariant Bethe-Salpeter wave functions along with the meson masses and leptonic decay constants listed in Table~\ref{params}\,B.  For the ground state pion, both RL and bRL predictions are a good match with empirical determinations \cite{ParticleDataGroup:2024cfk}.  This outcome is an implicit expression of the Nambu-Goldstone boson character of the pion \cite{Maris:1997hd, Maris:1997tm, Chang:2009zb, Fischer:2009jm, Qin:2014vya}: all sym\-metry-preserving CSM truncations must deliver a good description.
Regarding $\pi_{s\bar s}$, the meson masses are the same, but one sees a $15$\% difference between the decay constants.
This is an indication of the observable impact of bRL corrections in the neighbourhood of the $s$ quark mass, \textit{viz}.\ following this size of Higgs-boson induced shift away from the Nambu-Goldstone boson limit.

Widespread use has shown that when the product $\omega D$ is kept fixed, one typically finds that ground-state pseudoscalar meson observables remain practically unchanged under $\omega \to (1\pm 0.1)\omega$ \cite{Qin:2020rad}.
Thus, since the quark current-masses are fixed by the meson masses, then $\omega$ is the only real parameter in Table~\ref{params}\,A.
It is fixed by requiring a realistic value of the $\pi$ leptonic decay constant.

\section{LFWF: formulae}
\label{Sec3}
The Poincar\'e-covariant Bethe-Salpeter wave function for a pseudoscalar meson can be written in the form
\begin{align}
& {\mathpzc X}_{\mathsf 5}(k;P)  =
S(k_\eta) \big\{ \gamma_5 \big[
i {E}_{\mathsf 5}(k;p) + k\cdot P {F}_{\mathsf 5}(k;P) \nonumber \\
&
\; + k\cdot P \gamma\cdot k {G}_{\mathsf 5}(k;p) + \sigma_{\mu\nu} k_\mu P_\nu {H}_{\mathsf 5}(k;P) \big]\big\}S(k_{\bar\eta})\,,
\label{BSWF}
\end{align}
where $S$ is the $2$-point Schwinger function (propagator) associated with the valence degrees of freedom in the bound-state;
$P$ is the meson total momentum, $P^2 = -m_{\mathsf 5}^2$, with $m_{\mathsf 5}$ the meson mass;
$k$ is the relative momentum between the valence quark and antiquark whose properties specify the character of the system, and $k_\eta = k+\eta P$, $k_{\bar\eta} = k-(1-\eta) P$, $0\leq \eta \leq 1$.
For mesons built from mass-degenerate valence degrees of freedom, it is useful to choose $\eta=1/2$; then, $E, F, G, H$ in Eq.\,\eqref{BSWF} are even functions of $k\cdot P$.
Projected into the rest-frame, the wave function in Eq.\,\eqref{BSWF} describes a meson with both $\mathsf S$- and $\mathsf P$-wave orbital angular momentum components \cite{Bhagwat:2006xi, Krassnigg:2009zh}.
In all subsequent calculations, the canonically normalised Bethe-Salpeter wave function is used; see, \textit{e.g}., Ref.\,\cite[Sec.\,3]{Nakanishi:1969ph} for discussion of this normalisation definition.

A pseudoscalar meson LFWF has two components:
\begin{align}
\psi_{\mathsf 5}^{\mathpzc L}(x,k_\perp^2)
& \propto
\gamma_5 \big[\gamma\cdot n \,\psi_{\mathsf 5}^{0}(x,k_\perp^2)\nonumber \\
& \qquad + i \sigma_{\mu\nu}n_\mu k_{\perp \nu} \, \psi_{\mathsf 5}^{1}(x,k_\perp^2)\big]\,,
\label{EqLFWF}
\end{align}
where $n$ is a lightlike $4$-vector, $n^2=0$, $n\cdot k_\perp = 0$,
$x=n\cdot k_\eta/n\cdot P$,
and $n\cdot P=-m_{\mathsf 5}$ in the meson rest frame.
The superscript on $\psi$ indicates the light-front orbital angular momentum (helicity) projection: the ${\mathpzc L}=0=\uparrow\downarrow=\downarrow\uparrow$ wave function has the light-front spins of the valence constituents antialigned; and ${\mathpzc L}=1=\uparrow\uparrow=\downarrow\downarrow$ has them aligned.
In our conventions, the $\mathpzc L=0$ component has mass dimension $1/{\rm GeV}$ and that of $\mathpzc L=1$ is $1/{\rm GeV}^2$.

Working with Eq.\,\eqref{BSWF}, the independent components in Eq.\,\eqref{EqLFWF} may be obtained via light-front projections:
\begin{equation}
    \psi_{\mathsf 5}^{\mathpzc L}(x,k_\perp^2)
     = {\rm tr}_{\rm CD}\int\frac{dk_3 dk_4}{\pi}
    \delta(x n\cdot P - n\cdot k_\eta)
    {\mathpzc P}^{\mathpzc L} {\mathpzc X}_{\mathsf 5}(k;P)\,,
    \label{LFWFP}
\end{equation}
where the trace is over colour and spinor indices
and
\begin{equation}
{\mathpzc P}^{0} = \tfrac{1}{4}\gamma_5\gamma\cdot n \,,
\quad {\mathpzc P}^{1} = \tfrac{1}{4}\tfrac{i}{k_\perp^2} \gamma_5 \sigma_{\mu\nu} n_\mu k_{\perp\nu}\,.
\end{equation}

The leading-twist two quasiparticle distribution amplitude (DA), calculated in Ref.\,\cite{Chang:2013pq}, is straightforwardly obtained from Eq.\,\eqref{LFWFP}:
\begin{equation}
f_{\mathsf 5} \varphi_{\mathsf 5}(x) =
\int^\Lambda \frac{d^2k_\perp}{ 16\pi^3}
 \psi_{\mathsf 5}^{0}(x,k_\perp^2)\,,
 \label{DefineDA}
\end{equation}
where $f_{\mathsf 5}$ is the meson's leptonic decay constant and
$\int^\Lambda$ indicates that a translationally invariant regularisation scheme must be used to calculate what is now effectively a four-dimensional integral.  The Mellin moments of this DA are plainly $k_\perp^2$-independent.

The associated pseudoscalar meson helicity-in\-de\-pen\-dent TMD is even more straightforward:
\begin{equation}
f_1(x,k_\perp^2) =
\frac{1}{(2\pi)^3}
\left[ | \psi_{\mathsf 5}^{0}(x,k_\perp^2)|^2
+ k_\perp^2 | \psi_{\mathsf 5}^{1}(x,k_\perp^2)|^2\right] \,,
\label{DefTMD}
\end{equation}
from which follows the pseudoscalar meson valence quasiparticle distribution function:
\begin{equation}
{\mathpzc q}_{\mathsf 5}(x) =
\int d^2k\, f_1(x,k_\perp^2) \,.
\label{DefDF}
\end{equation}
Since we are dealing with ``isospin''-symmetric states, there is no need to distinguish between particle and antiparticle valence constituents.  Normalisation of the LFWF is equivalent to baryon number conservation:
\begin{equation}
    \int_0^1 dx\, {\mathpzc q}_{\mathsf 5}(x) = 1\,.
    \label{EqBNcons}
\end{equation}

\begin{figure*}[t]
\hspace*{-1ex}\begin{tabular}{lcl}
\large{\textsf{A}$_{\rm RL}^{\pi\uparrow\downarrow}$} & & \large{\textsf{B}$_{\rm RL}^{\pi\uparrow\uparrow}$}\\[-0.5ex]
%
\includegraphics[clip, width=0.45\textwidth]{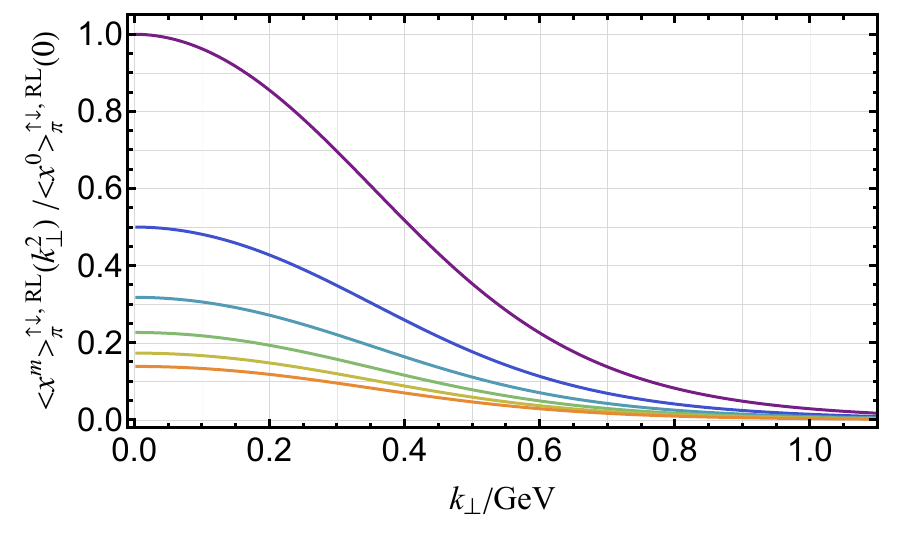} & \hspace*{-0.5em} &
\includegraphics[clip, width=0.45\textwidth]{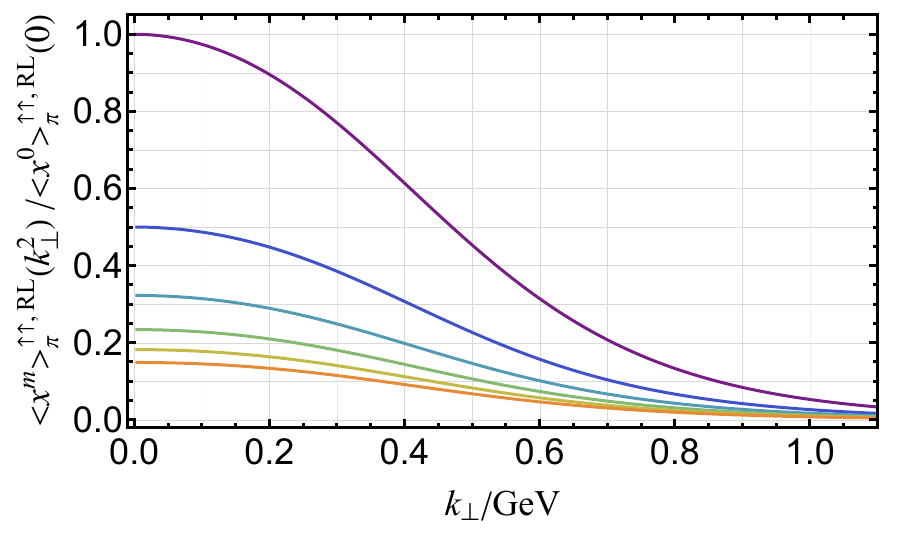}
\end{tabular}
\hspace*{-1ex}\begin{tabular}{lcl}
\large{\textsf{C}$_{\rm bRL}^{\pi\uparrow\downarrow}$} & & \large{\textsf{D}$_{\rm bRL}^{\pi\uparrow\uparrow}$}\\[0.3ex]
%
\includegraphics[clip, width=0.45\textwidth]{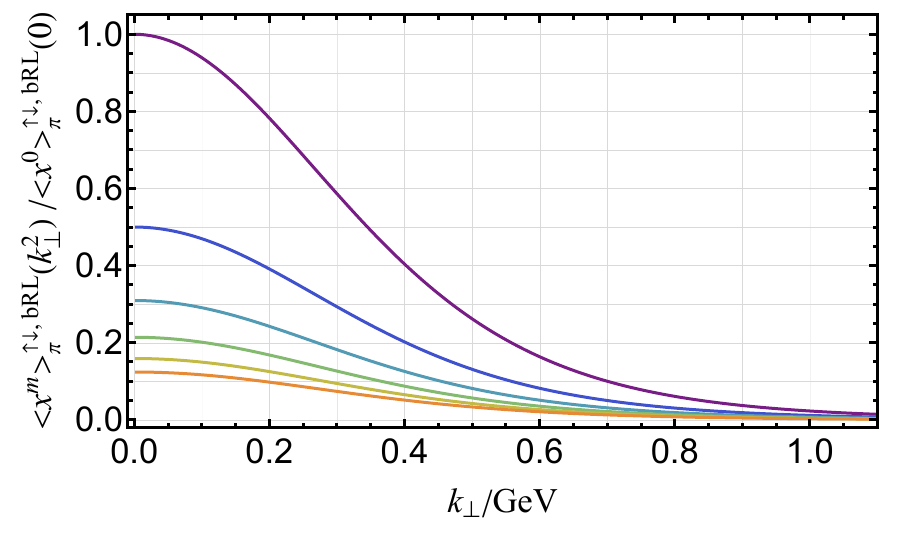} & \hspace*{-0.5em} &
\includegraphics[clip, width=0.45\textwidth]{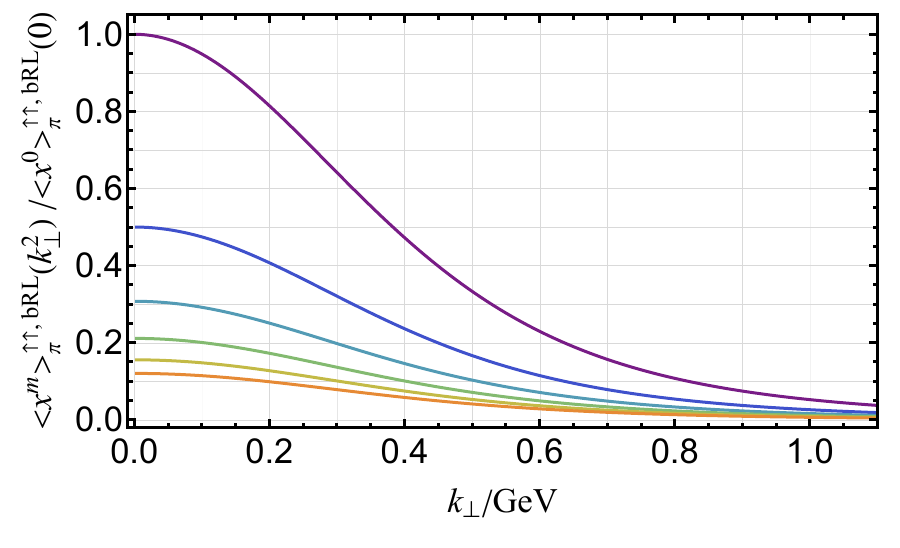}
\end{tabular}
\caption{\label{Fmoments}
Mellin moments, $m=0,\ldots, 5$, of $\pi$ LFWF, defined by Eq.\,\eqref{PionMMs}: the magnitudes decrease with increasing $m$.
{\sf Panel A}.  RL, spins antialigned,
$\langle x^0 \rangle_\pi^{\uparrow\downarrow}(0)=55.9/\Lambda_I$.
{\sf Panel B}.  RL, spins aligned,
$\langle x^0 \rangle_\pi^{\uparrow\uparrow}(0)=(9.61/\Lambda_I)^2$.
{\sf Panel C}.  bRL, spins antialigned,
$\langle x^0 \rangle_\pi^{\uparrow\downarrow}(0)=72.5/\Lambda_I$.
{\sf Panel D}.  bRL, spins aligned.
$\langle x^0 \rangle_\pi^{\uparrow\uparrow}(0)=(12.8/\Lambda_I)^2$.
(Recall $\Lambda_I=1\,$GeV, Eq.\,\eqref{defcalG}.)
}
\end{figure*}

Here, it is worth reiterating that we deliver results at the hadron scale, $\zeta_{\cal H}$.
This scale is defined in the context of QCD effective charges \cite{Grunberg:1980ja, Grunberg:1982fw}, wherewith a QCD running coupling is defined using the expansion of a chosen observable restricted to first order in the perturbative coupling, $\alpha_s$; see the discussion in Ref.\,\cite[Sec.\,4.3]{Deur:2023dzc}.
It follows that effective charges are typically process dependent, \textit{e.g}., $\alpha_{g_1}(k^2)$, defined via the Bjorken sum rule \cite{Deur:2022msf}.
Any such coupling is analogous to the Gell-Mann--Low coupling in quantum electrodynamics \cite{GellMann:1954fq} and has the following properties: the coupling is consistent with the QCD renormalisation group; renormalisation scheme independent; everywhere analytic and finite; and provides an infrared completion of any per\-tur\-batively-defined (standard) running coupling.

\begin{figure*}[t]
\hspace*{-1ex}\begin{tabular}{lcl}
%
%
\large{\textsf{A}$_{\rm RL}^{\pi\uparrow\downarrow}$} & & \large{\textsf{B}$_{\rm RL}^{\pi\uparrow\uparrow}$}\\[-0.5ex]
%
\includegraphics[clip, width=0.45\textwidth]{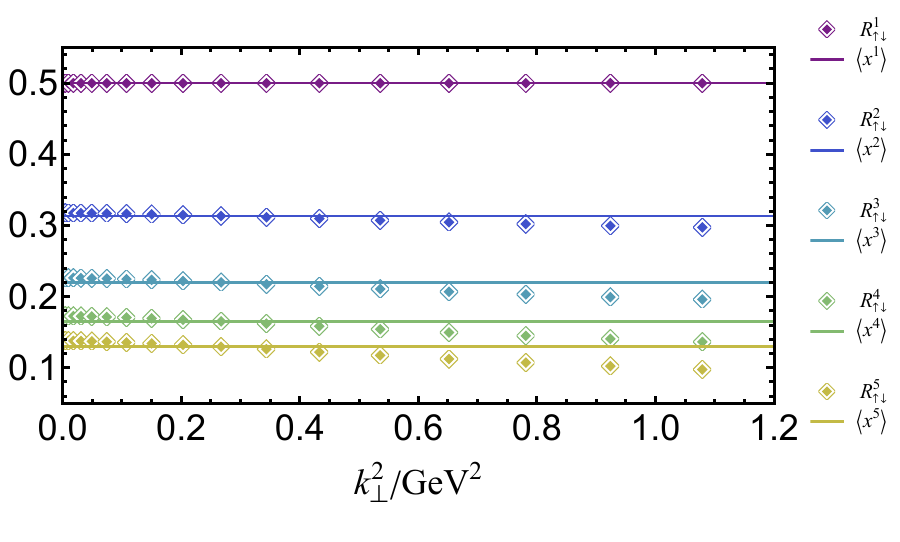} & \hspace*{-0.5em} &
\includegraphics[clip, width=0.45\textwidth]{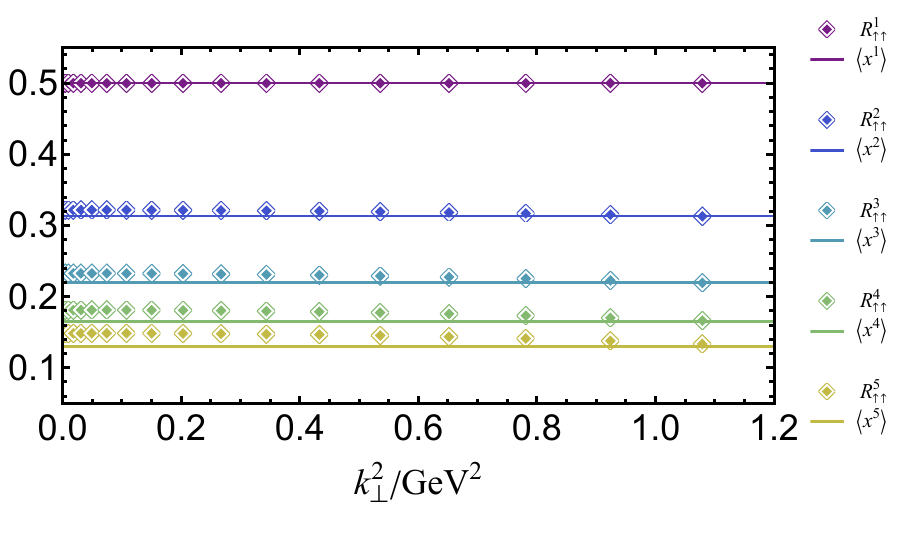}
\end{tabular}
\hspace*{-1ex}\begin{tabular}{lcl}
%
%
\large{\textsf{C}$_{\rm bRL}^{\pi\uparrow\downarrow}$} & & \large{\textsf{D}$_{\rm bRL}^{\pi\uparrow\uparrow}$}\\[0.3ex]
%
\includegraphics[clip, width=0.45\textwidth]{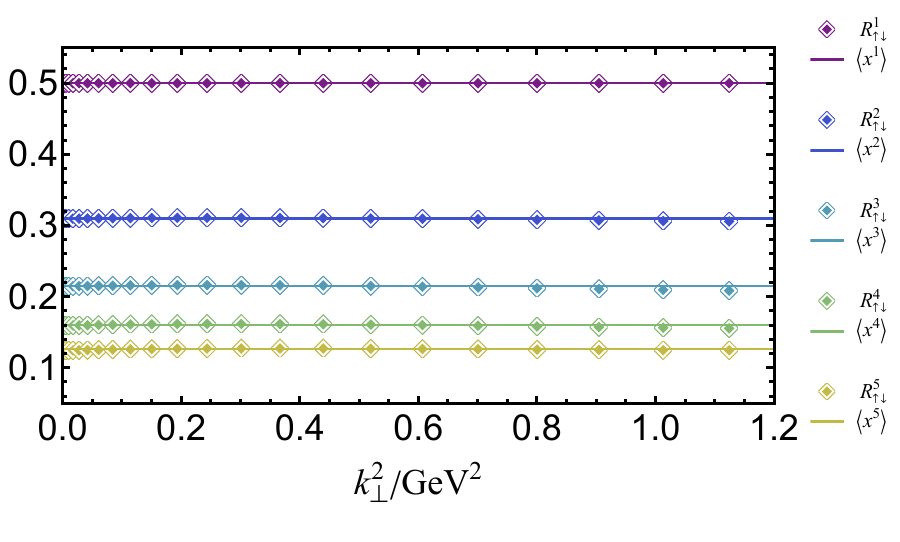} & \hspace*{-0.5em} &
\includegraphics[clip, width=0.45\textwidth]{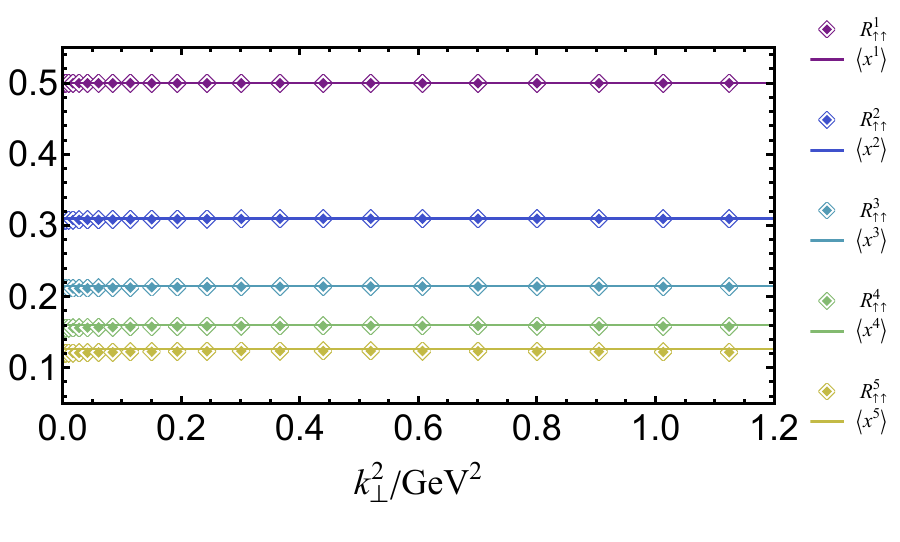}
\end{tabular}
\caption{\label{FmomentsSep}
Ratios of LFWF Mellin moments, Eq.\,\eqref{eq:ratios}.
In each panel, the solid lines mark the $k_\perp^2$-independent $m$ moment of the DA in Eq.\,\eqref{DefineDA}.
{\sf Panel A}.  RL, spins antialigned.
{\sf Panel B}.  RL, spins aligned.
{\sf Panel C}.  bRL, spins antialigned.
{\sf Panel D}.  bRL, spins aligned.
}
\end{figure*}

We adopt the effective charge elucidated in Ref.\,\cite{Yin:2023dbw}, denoted $\alpha_{1\ell}(k^2)$.
This is a function which, when used to integrate the leading-order perturbative DGLAP equations \cite{Dokshitzer:1977sg, Gribov:1971zn, Lipatov:1974qm, Altarelli:1977zs}, defines an evolution scheme for all parton distribution functions (DFs) and for any hadron that is all-orders exact.
This is the all-orders (AO) approach.
Its definition is broader than usual because it refers to an entire class of observables, not just a single measurable quantity.
We stress that the pointwise form of $\alpha_{1\ell}(k^2)$ is largely irrelevant given that a large number of model-independent results can be proved without any reference to the $k^2$-profile of $\alpha_{1\ell}(k^2)$; see, \textit{e.g}., Refs.\,\cite{Cui:2022bxn, Xu:2023bwv, Lu:2023yna, Xu:2024nzp, Yao:2024ixu}.

Hereafter we exploit the fact that, with an effective charge defined in this way, there is a resolving scale, $\zeta=\zeta_{\cal H} < m_N$, at which all properties of a given hadron are carried by its dressed valence degrees of freedom.
Consequently, the LFWFs and TMDs determined herein are expressed with reference to a quasiparticle Fock space defined by a light-front Hamiltonian that corresponds to the CSM truncations described above.  Since the LFWFs are obtained by light-front projection of a given Poincar\'e-covariant Bethe-Salpeter wave function, then the Hamiltonian need not be specified.  It is sufficient to record here that the leading component in the quasiparticle Fock space consists of an enumerable infinity of gluon, quark, and antiquark parton Fock space vectors.

\section{LFWF: results}
\label{Sec4}
Direct evaluation of the integrals specified by Eq\,\eqref{LFWFP} would require detailed knowledge of ${\mathpzc X}_{\mathsf 5}(k;P)$ in the complex-$k_
\eta^2$ plane.  That need can be circumvented by working with the following $k_\perp^2$-dependent Mellin moments:
{\allowdisplaybreaks
\begin{subequations}
\label{PionMMs}
\begin{align}
& \langle x^m  \rangle^{\mathpzc L} (k_\perp^2)
 = \int_0^1 dx\,x^m \, \psi_{\mathsf 5}^{\mathpzc L}(x,k_\perp^2) \\
& = \frac{1}{n\cdot P }{\rm tr}_{\rm CD}\int\frac{dk_3 dk_4}{\pi}
\, \left[\frac{n\cdot k_\eta}{n\cdot P}\right]^m
{\mathpzc P}^{\mathpzc L} {\mathpzc X}_{\mathsf 5}(k;P)\,.
\end{align}
\end{subequations}}

We calculate these moments using the approach discussed in connection with Ref.\,\cite[Eq.\,(15)]{Xu:2025cyj}; and, furthermore, reexpress the measure and integrand in polar coordinates.  The latter step works to improve the precision of interpolations used in the numerical integration.
Following this prescription, stable results for the $k_\perp^2$-dependence of moments $m=0,1,\ldots,5$ are readily obtained.

In reconstructing a LFWF from this set of moments,
a systematic uncertainty is introduced by neglecting $m\geq 6$; but it is small so long as flexible, physically constrained reconstruction functions are chosen; see, \textit{e.g}., Refs.\,\cite{Chang:2013pq, Cloet:2013tta, Cui:2020tdf, Cui:2022bxn, Xu:2024nzp}.


The calculated moments are drawn in Fig.\,\ref{Fmoments}.
Comparing upper with lower panels, one sees that the RL LFWF moments exhibit a slower fall-off with $k_\perp^2$ than the bRL moments.
Hence, the RL LFWF must be broader in $k_\perp$-space or, equally, more pointlike in the conjugate configuration space.
It follows, moreover, that the RL DA, Eq.\,\eqref{DefineDA}, must also be broader than the bRL DA.  This result was first reported in Ref.\,\cite{Chang:2013pq}.
The conclusion is supported by numerical simulations of lattice-regularised QCD (lQCD) \cite{RQCD:2019osh}.

In Fig.\,\ref{FmomentsSep}, we plot the following dimensionless ratios:
\begin{align}
\label{eq:ratios}
R_\pi^{m,\, \mathpzc{L}}(k_\perp^2) =
\frac{\langle x^m \rangle_\pi^{\mathpzc L}(k_\perp^2)}{\langle x^0 \rangle_\pi^{\mathpzc L}(k_\perp^2)}.
\end{align}
Considering the RL results, Figs.\,\ref{FmomentsSep}\,A, B, one sees that the ratio is $k_\perp^2$-independent to a fair degree
and the value for each moment is close to that of the analogous moment obtained from the DA in Eq.\,\eqref{DefineDA}.
Furthermore, the ratios for both spin projections are practically equivalent.
One can therefore conclude that the RL pion LFWF -- both spin components -- can approximately be written in a factorised form:
\begin{equation}
\label{EqSep}
\psi_{\mathsf 5}^{\mathpzc L}(x,k_\perp^2)
\approx \varphi_{\mathsf 5}(x) \times
F^{\mathpzc L}(k_\perp^2)\,.
\end{equation}

Importantly, the images indicate that both spin projections have the same $x$-dependence, \textit{i.e}., are characterised by the same DA; see, also, \ref{AppFit}\,--\,Table~\ref{tab:momentspi}.
As explained and exploited, \textit{e.g}., in Refs.\,\cite[Sec.\,3]{Roberts:2021nhw}, \cite{Cui:2020tdf}, \cite{Raya:2021zrz}, it follows that Eq.\,\eqref{DefDF} entails
\begin{equation}
    {\mathpzc q}_{\mathsf 5}(x) \propto \varphi_{\mathsf 5}^2(x)\,,
\end{equation}
notwithstanding the presence of the LFWF $\mathpzc L=1$ component.

On the other hand, analyses in perturbative QCD predict \cite{Lepage:1980fj}:
\begin{equation}
    k^2 \gg m_N^2 \, | \quad
    \psi_{\mathsf 5}^{0}(x,k_\perp^2) \propto \frac{1}{k_\perp^2}
    \propto k_\perp^2 \psi_{\mathsf 5}^{1}(x,k_\perp^2)\,,
    \label{LFWFUV}
\end{equation}
up to (damping) $\ln k_\perp^2$ corrections.
CSM predictions, including those obtained and discussed herein, are consistent with this behaviour; see, \emph{e.g}., Refs.\,\cite{Ding:2018xwy, Yao:2024drm}, which demonstrate that relevant perturbative-QCD hard-scat\-tering formulae \cite{Lepage:1980fj} are preserved.
Consequently, realistic representations of LFWF $k_\perp^2$ dependence must express different ultraviolet behaviour for the independent spin projections.

Turning to ratio results obtained with the EHM-improved bRL kernel, Figs.\,\ref{FmomentsSep}\,C, D, one sees that what were approximately $k_\perp^2$-independent profiles in RL truncation, which were reasonably well matched with the associated DA moment, become constant profiles with $k_\perp^2$-independent values that are practically indistinguishable from the linked DA moment.  Thus whilst Eq.\,\eqref{EqSep} provides a fair approximation to the RL results, it delivers a reliable pointwise representation for the bRL pion LFWF.
(See, also, \ref{AppFit}\,--\,Table~\ref{tab:momentspi}.)

Qualitatively equivalent statements can be made about the $\pi_{s \bar s}$ moments and LFWFs.

The results described above provide strong support for the approach to calculating pion and kaon DFs and generalised parton distributions in, \textit{e.g}.,  Refs.\,\cite{Cui:2020tdf, Raya:2021zrz}.

\begin{table}[t!]
\centering
\caption{%
Values of LFWF coefficients in Eq.\,\eqref{RealSep} for the various truncations and systems considered herein.
\textit{N.B}.\ In all cases, within mutual uncertainties, $\rho^{(0=\uparrow\downarrow)}=\rho^{(1=\uparrow\uparrow)}$.
Associated correlation matrices are collected in \ref{AppFit}.
\label{tab:paramspi}}
\begin{tabular*}
{\hsize}
{
l@{\extracolsep{0ptplus1fil}}
cl@{\extracolsep{0ptplus1fil}}
c@{\extracolsep{0ptplus1fil}}
c@{\extracolsep{0ptplus1fil}}
c@{\extracolsep{0ptplus1fil}}}\hline\hline
\centering
${\mathpzc p}_{{\mathsf 5}R}^{\mathpzc L}$ & $a_{0}^{{\mathpzc L}}$ & $b_{1}^{{\mathpzc L}}$ & $b_{2}^{{\mathpzc L}}$& $\rho^{({\mathpzc L})}$ \\
\hline
${\psi}^{\uparrow\downarrow}_{\pi,\,\mathrm{RL}}$&
$\phantom{1}51.21(2.85)$ & $1.83(60)$&$20.45(1.18)$&$0.086(12)$\\
${\psi}^{\uparrow\uparrow}_{\pi,\,\mathrm{RL}}$
&$\phantom{1}91.98(0.24)$&$1.99(09)$&$12.21(0.34)$&$0.031(04)$ \\
${\psi}^{\uparrow\downarrow}_{\pi,\,\mathrm{bRL}}$&
$\phantom{1}67.76(3.13)$ & $4.41(63)$&$26.14(1.35)$&$0.224(19)$ \\
${\psi}^{\uparrow\uparrow}_{\pi,\,\mathrm{bRL}}$&
$164.30(0.24)$&$5.04(07)$&$12.20(0.26)$&$0.285(11)$ \\
\hline
${\psi}^{\uparrow\downarrow}_{\pi_{s \bar s},\,\mathrm{RL}}$&$\phantom{1}39.98(2.52)$&$0.78(41)$&$\phantom{2}8.31(0.52)$&
$\rho_{\rm as}$\\
${\psi}^{\uparrow\uparrow}_{{\pi_{s \bar s},\,\mathrm{RL}}}$&
$\phantom{1}49.39(0.23)$&$1.11(10)$&$\phantom{2}5.46(0.25)$&
$\rho_{\rm as}$
\\
${\psi}^{\uparrow\downarrow}_{{\pi_{s \bar s},\,\mathrm{bRL}}}$
&$\phantom{1}70.47(3.36)$&$3.56(57)$&$20.70(1.10)$&
$\rho_{\rm as}$\\
${\psi}^{\uparrow\uparrow}_{{\pi_{s \bar s},\,\mathrm{bRL}}}$
&$136.29(0.25)$&$4.12(08)$&$11.72(0.31)$ &
$\rho_{\rm as}$
\\
\hline\hline
\end{tabular*}

\medskip

\begin{tabular*}
{\hsize}
{
l@{\extracolsep{0ptplus1fil}}
c@{\extracolsep{0ptplus1fil}}
c@{\extracolsep{0ptplus1fil}}}\hline\hline
${\mathpzc p}_{{\mathsf 5}V}^{\mathpzc L}$ & $c_{0}^{{\mathpzc L}}$ & $d_{1}^{{\mathpzc L}}$ \\\hline
${\psi}^{\uparrow\downarrow}_{\pi,\,\mathrm{RL}}$&
$4.87(2.91)$&$48.24(29.98)\ $\\
${\psi}^{\uparrow\downarrow}_{\pi,\,\mathrm{bRL}}$&
$4.89(3.18)$ & $60.80(37.30)\ $ \\
${\psi}^{\uparrow\downarrow}_{\pi_{s \bar s},\,\mathrm{RL}}$ &$4.77(2.57)$&$30.45(17.05)\ $\\
${\psi}^{\uparrow\downarrow}_{{\pi_{s \bar s},\,\mathrm{bRL}}}$
&$5.34(3.42)$&$52.50(31.90)\ $\\
\hline\hline
\end{tabular*}

\end{table}


\begin{figure*}[t]
\hspace*{-1ex}\begin{tabular}{lcl}
\large{\textsf{A}$_{\rm RL}$} & & \large{\textsf{B}$_{\rm RL}$}\\[-0.7ex]
%
\includegraphics[clip, width=0.45\textwidth]{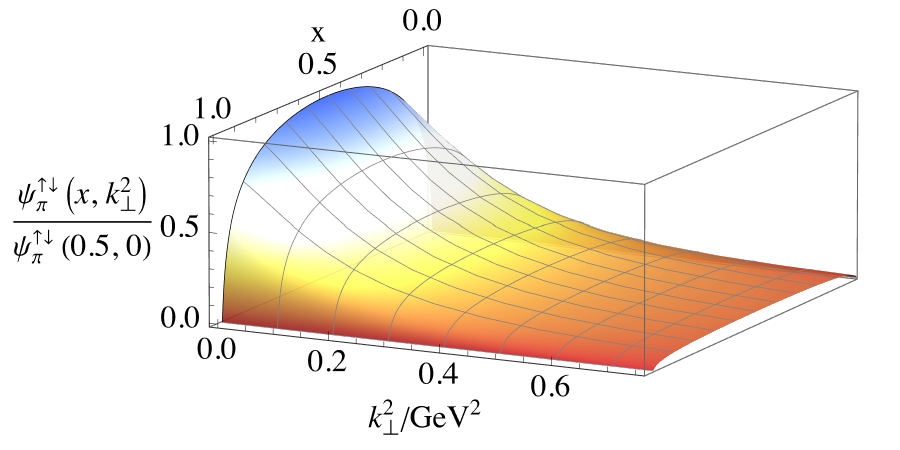} & \hspace*{-0.5em} &
\includegraphics[clip, width=0.45\textwidth]{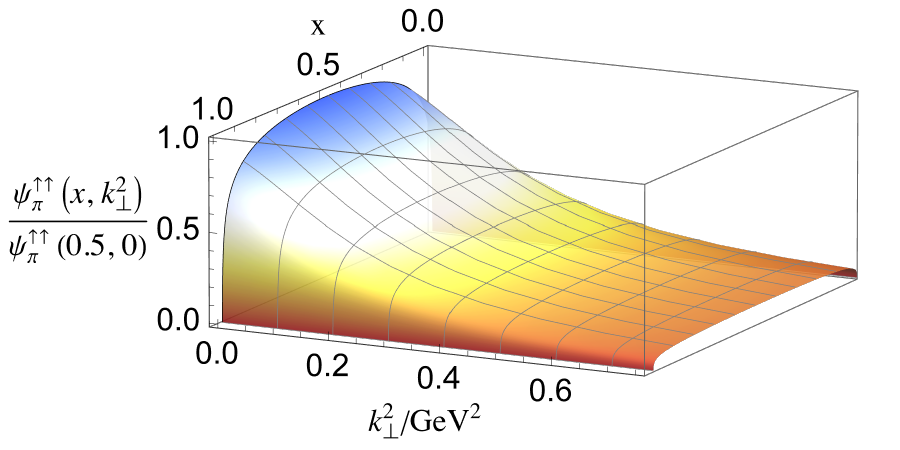}
\end{tabular}
\hspace*{-1ex}\begin{tabular}{lcl}
\large{\textsf{C}$_{\rm bRL}$} & & \large{\textsf{D}$_{\rm bRL}$}\\[-0.7ex]
%
\includegraphics[clip, width=0.45\textwidth]{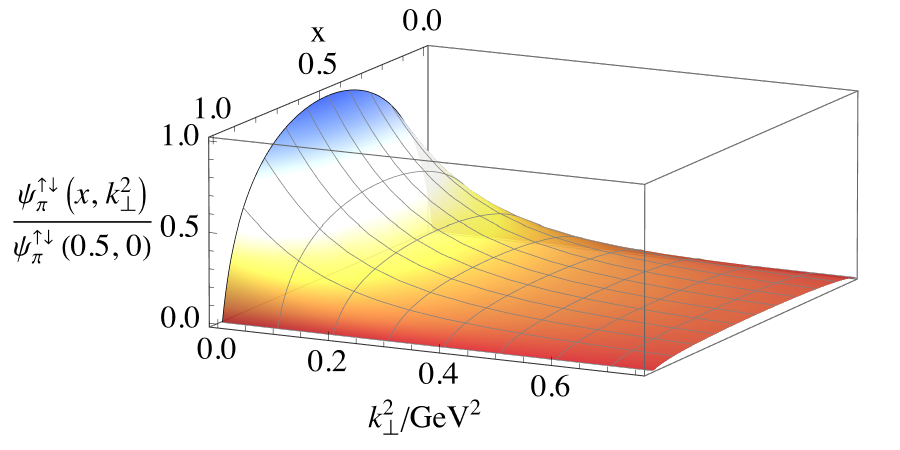} & \hspace*{-0.5em} &
\includegraphics[clip, width=0.45\textwidth]{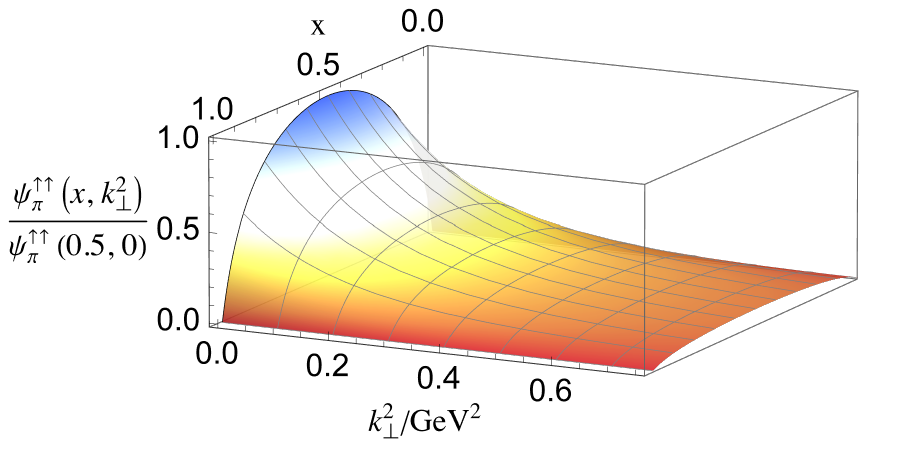}
\end{tabular}
%
\hspace*{-1ex}\begin{tabular}{lcl}
\large{\textsf{E}$_{\rm RL}$} & & \large{\textsf{F}$_{\rm RL}$}\\[-0.7ex]
%
\includegraphics[clip, width=0.45\textwidth]{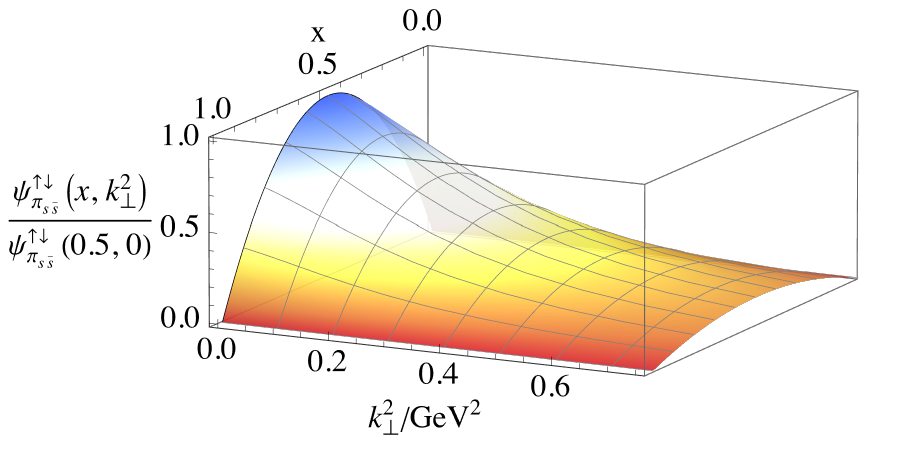} & \hspace*{-0.5em} &
\includegraphics[clip, width=0.45\textwidth]{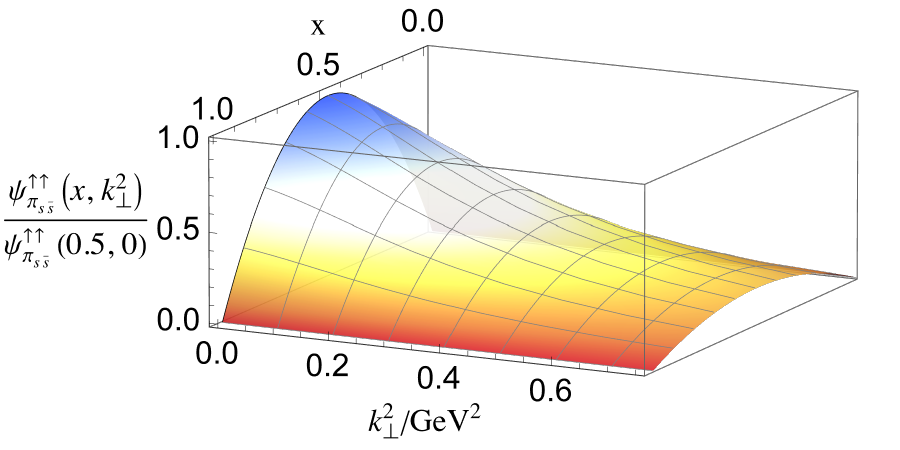}
\end{tabular}
\hspace*{-1ex}\begin{tabular}{lcl}
\large{\textsf{G}$_{\rm bRL}$} & & \large{\textsf{H}$_{\rm bRL}$}\\[-0.7ex]
%
\includegraphics[clip, width=0.45\textwidth]{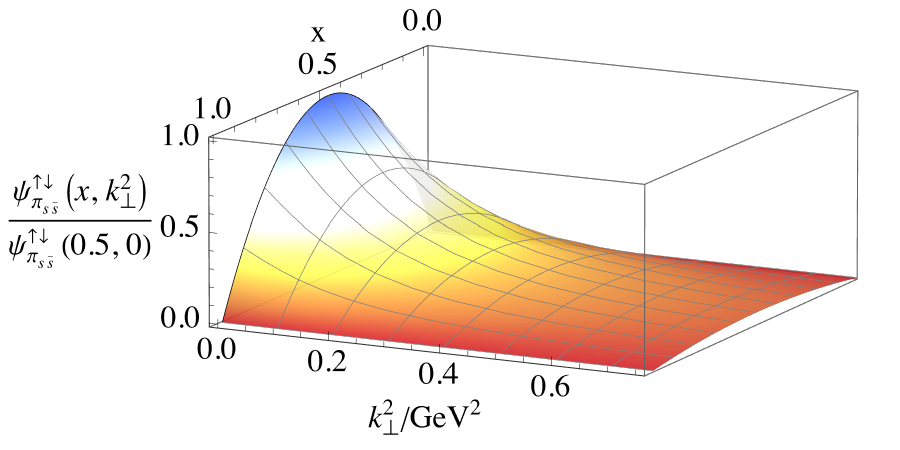} & \hspace*{-0.5em} &
\includegraphics[clip, width=0.45\textwidth]{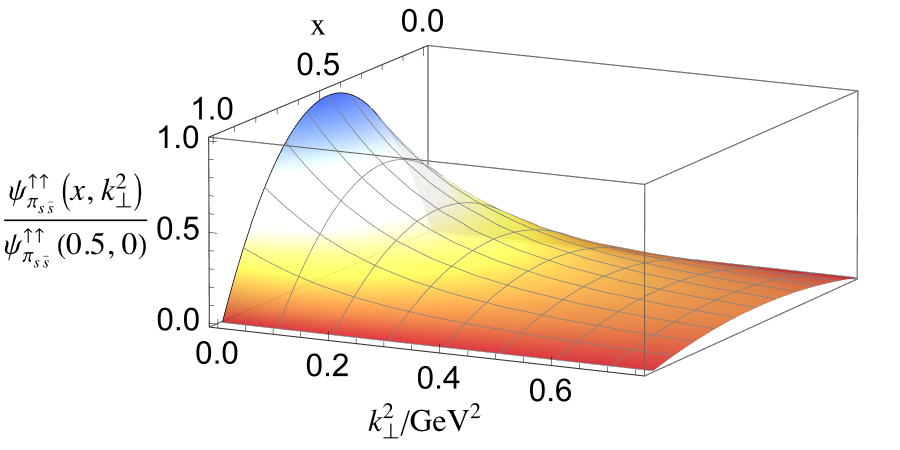}
\end{tabular}
\caption{\label{fig:LFWF}
LFWFs reconstructed from their Mellin moments, expressed by Eq.\,\eqref{RealSep} and central values of the coefficients in Table~\ref{tab:paramspi}.  Each curve is normalised by its peak size, determined by the associated values of $\rho$, $a_{0}^{{\mathpzc L}}+c_{0}^{{\mathpzc L}}$ in Table~\ref{tab:paramspi}.
{\sf Panel A}.  $\pi$ RL, spins antialigned.
{\sf Panel B}.  $\pi$ RL, spins aligned.
{\sf Panel C}.  $\pi$ bRL, spins antialigned.
{\sf Panel D}.  $\pi$ bRL, spins aligned.
{\sf Panel E}.  $\pi_{s\bar s}$ RL, spins antialigned.
{\sf Panel F}.  $\pi_{s\bar s}$ RL, spins aligned.
{\sf Panel G}.  $\pi_{s\bar s}$ bRL, spins antialigned.
{\sf Panel H}.  $\pi_{s\bar s}$ bRL, spins aligned.
}
\end{figure*}

{\allowdisplaybreaks
Having established that separable \textit{Ans\"atze} provide a sound representation of $\pi$, $\pi_{s \bar s}$ LFWFs, we introduce a particular form of Eq.\,\eqref{EqSep}, \textit{viz}.
\begin{equation}
\label{RealSep}
\psi_{\mathsf 5}^{\mathpzc L}(x,k_\perp^2)
=  \varphi_{\mathsf 5}(x) \times
{\mathpzc p}_{\mathsf 5}^{\mathpzc L}(k_\perp^2)\,,
\end{equation}
with
\begin{equation}
\varphi_{\mathsf 5}(x) = \tfrac{1}{{\mathsf N}}\ln [ 1+x(1-x)/\rho_{\mathsf 5}^2]\,,
\label{LogForm}
\end{equation}
where $\rho_{\mathsf 5}$ is a constant,
${\mathsf N}$ ensures $ \int_0^1 dx \varphi_{\mathsf 5}(x) =1$, consistent with Eq.\,\eqref{DefineDA}, and
\begin{subequations}
\label{Fitp}
\begin{align}
{\mathpzc p}_{\mathsf 5}^{\mathpzc L}(k_\perp^2)
& =
{\mathpzc p}_{{\mathsf 5}R}^{\mathpzc L}(k_\perp^2)
+ {\mathpzc p}_{{\mathsf 5}V}^{\mathpzc L}(k_\perp^2) \,,\\
{\mathpzc p}_{{\mathsf 5}R}^{\mathpzc L}(k_\perp^2)
& = \frac{a_0^{\mathpzc L} /\Lambda_I^{1+{\mathpzc L}}}
{1+b_1^{\mathpzc L}k_\perp^2/\Lambda_I^2 + b_2^{\mathpzc L} k_\perp^4/\Lambda_I^4}\,,\\
{\mathpzc p}_{{\mathsf 5}V}^{\mathpzc L}(k_\perp^2)
& = \frac{c_0^{\mathpzc L} /\Lambda_I^{1+{\mathpzc L}}}
{1+d_1^{\mathpzc L}k_\perp^2/\Lambda_I^2 }\,.
\end{align}
\end{subequations}
Notably, for any $\rho_{\mathsf 5} \gtrsim 1$ in Eq.\,\eqref{LogForm}, the $x$-profile of $\psi_{\mathsf 5}^{\mathpzc L}(x,k_\perp^2)$ is essentially equivalent to that of the QCD asymptotic DA \cite{Lepage:1979zb, Efremov:1979qk, Lepage:1980fj}: $\varphi_{\rm as}(x) = 6 x(1-x)$.
Indeed, using the ${\mathsf L}_1$ measure, the curves obtained with $\rho_5=1, 2$ differ by just 1.4\%, with $\rho_5=2$ differing from $\varphi_{\rm as}$ by just 0.52\%.
Moreover, in order to ensure reconstruction consistency with Eq.\,\eqref{LFWFUV}, two independent Pad\'e approximants are used to express ${\mathpzc p}_{\mathsf 5}^{\mathpzc L}(k_\perp^2)$.
For ${\mathpzc L}=0$, all five reconstruction parameters are nonzero in the $k_\perp^2$ term; but for ${\mathpzc L}=1$, $c_0^{{\mathpzc L}=1} \equiv 0 \Rightarrow {\mathpzc p}_{{\mathsf 5}V}^{\mathpzc L=1} \equiv 0$, \textit{viz}.\ there are only three parameters.
}

Using this product \textit{Ansatz} and the least-squares fitting procedure described in \ref{AppFit}, one arrives at the dimensionless best-fit reconstruction parameters listed in Table~\ref{tab:paramspi}.
The entry $\rho_{\rm as}$ means that, within any sensible understanding of the precision that may be achieved,  the $x$-profile is indistinguishable from $\varphi_{\rm as}$.  As noted above, any value $\rho \gtrsim 1$ is practically equivalent in this case.
The associated correlation matrices are listed in \ref{AppFit}.

Using the central values in Table~\ref{tab:paramspi}, one produces the LFWF images in Fig.\,\ref{fig:LFWF}.  The following points are worth stressing.
\begin{enumerate}[label=(\textit{\roman*})]
\item In each case, comparing kindred RL and bRL results, a given RL LFWF damps less rapidly as the probe is moved in any direction away from its global maximum, \textit{i.e}., RL LFWFs are (far) more dilated than bRL LFWFs. (Reference~\cite{Chang:2013pq} demonstrated this for DAs.)
Consequently, RL truncation produces a bound-state that is more compact in the conjugate coordinate space than the bRL kernel.
Available comparisons with inferences from data and other robust calculations indicate that bRL results are more realistic; see, \textit{e.g}., Refs.\,\cite{Cui:2020tdf, Cui:2021mom, Cui:2022bxn, Xu:2022kng, Xing:2023wuk, Roberts:2023lap, Xu:2025cyj}.
Thus the true extent of LFWFs is best measured by the bRL predictions.
%
\item With increased quark current mass, the $x$ dependence of each LFWF is less dilated.
The reduced dilation, \textit{i.e}., contraction, is an expression in the DA of Higgs-boson modulation of EHM.
In fact, as revealed in Refs.\,\cite{Ding:2015rkn, Chen:2018rwz, Ding:2018xwy, Roberts:2020udq}, confirmed via lQCD \cite{Zhang:2020gaj}, and reproduced by our analysis, the $\pi_{s \bar s}$ DA is practically indistinguishable from the QCD asymptotic DA \cite{Lepage:1979zb, Efremov:1979qk, Lepage:1980fj}: $\varphi_{\pi_{s\bar s}}\approx \varphi_{\rm as}$.

\item Regarding the LFWF $k_\perp^2$ dependence, the images in Fig.\,\ref{fig:LFWF} paint an ambiguous picture.
On the domain depicted, with increased quark current-mass, RL truncation obviously delivers results with dilated $k_\perp^2$ profiles.
In contrast, the $k_\perp^2$ profiles of the more realistic bRL LFWFs are fairly insensitive to the change in quark current mass.

\end{enumerate}

\begin{figure}[t]
\hspace*{-1ex}\begin{tabular}{l}
\large{\textsf{A}$_{\rm RL}$}\\[-0.3ex]
%
\includegraphics[clip, width=0.45\textwidth]{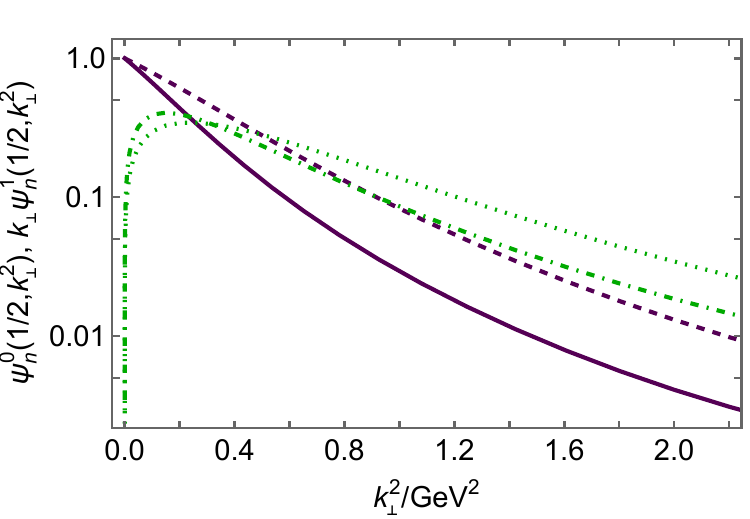}
\end{tabular}
\hspace*{-1ex}\begin{tabular}{l}
\large{\textsf{B}$_{\rm bRL}$} \\[0.3ex]
\includegraphics[clip, width=0.45\textwidth]{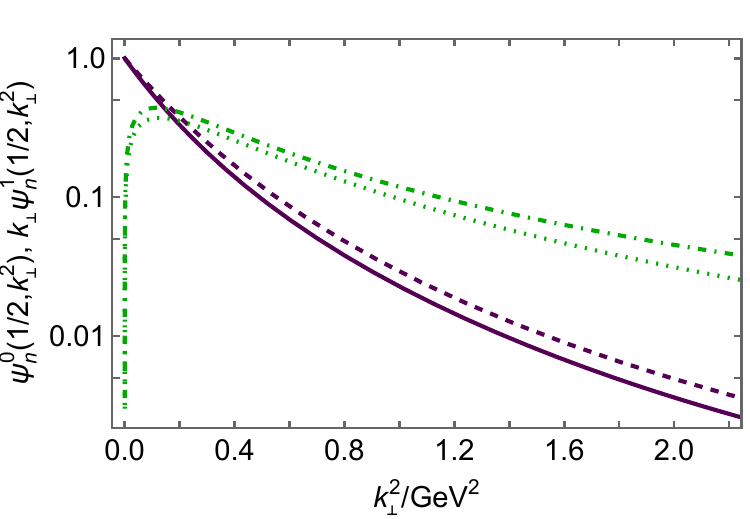}
\end{tabular}
\caption{\label{F4T}
LFWFs for the $\pi$ and $\pi_{s\bar s}$.
{\sf Panel A} -- RL truncation;
and
{\sf Panel B} -- bRL truncation.
Legend, both panels:
solid purple -- $\psi_\pi^0$ and dashed purple -- $\psi_{\pi_{s\bar s}}^0$; and
dot-dashed green -- $|k_\perp|\psi_\pi^1$ and dotted green -- $|k_\perp|\psi_{\pi_{s\bar s}}^1$.
Further, in each panel, the LFWFs are normalised by the appropriate value of $\psi^0(1/2,0)$, so the results are dimensionless and the $k_\perp^2$-dependence is directly comparable.  This is the meaning of the subscript $n$ on the ordinate labels.
}
\end{figure}

\begin{figure*}[t]
\hspace*{-1ex}\begin{tabular}{lcl}
\large{\textsf{A}$_{\rm RL}^\pi$} & & \large{\textsf{B}$_{\rm bRL}^{\pi}$}\\[-1ex]
%
\includegraphics[clip, width=0.45\textwidth]{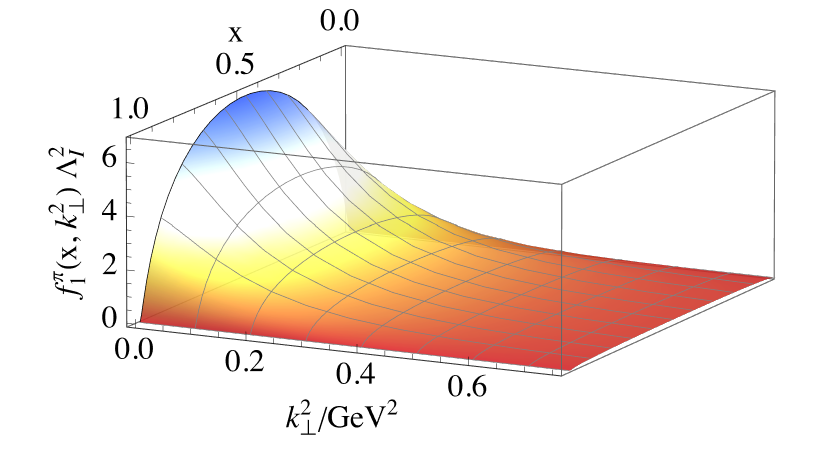} & \hspace*{-0.5em} &
\includegraphics[clip, width=0.45\textwidth]{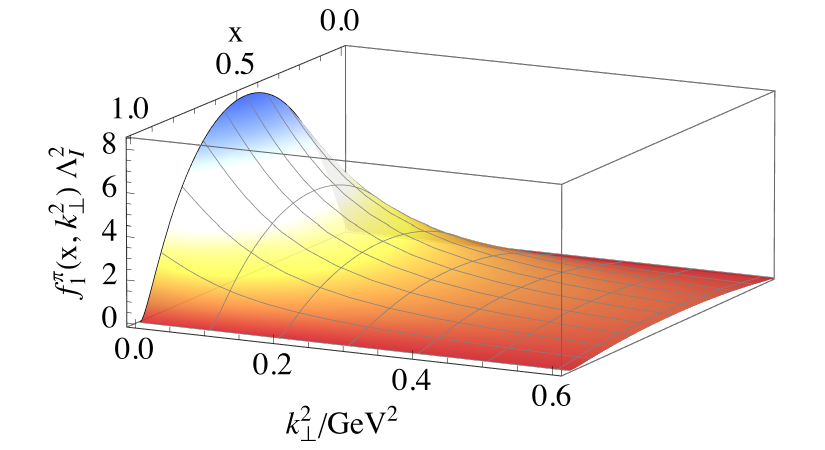}
\end{tabular}
\hspace*{-1ex}\begin{tabular}{lcl}
\large{\textsf{C$_{\rm RL}^{\pi_{s\bar s}}$}}
& & \large{\textsf{D}$_{\rm bRL}^{\pi_{s\bar s}}$}\\[-1ex]
%
\includegraphics[clip, width=0.45\textwidth]{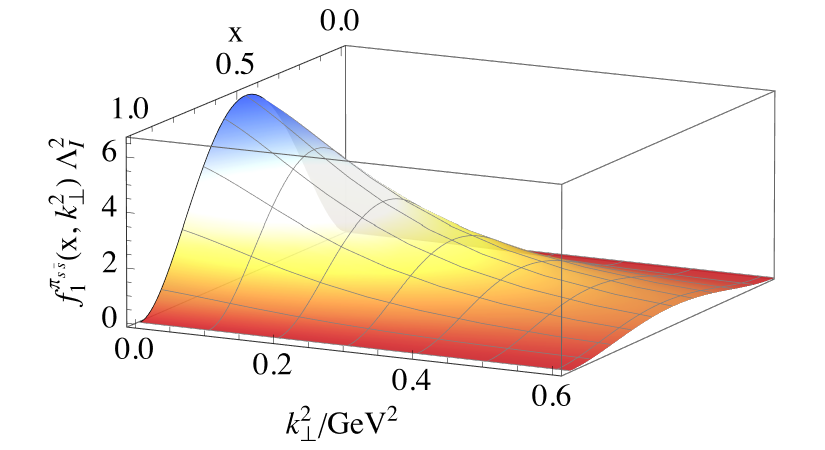} & \hspace*{-0.5em} &
\includegraphics[clip, width=0.45\textwidth]{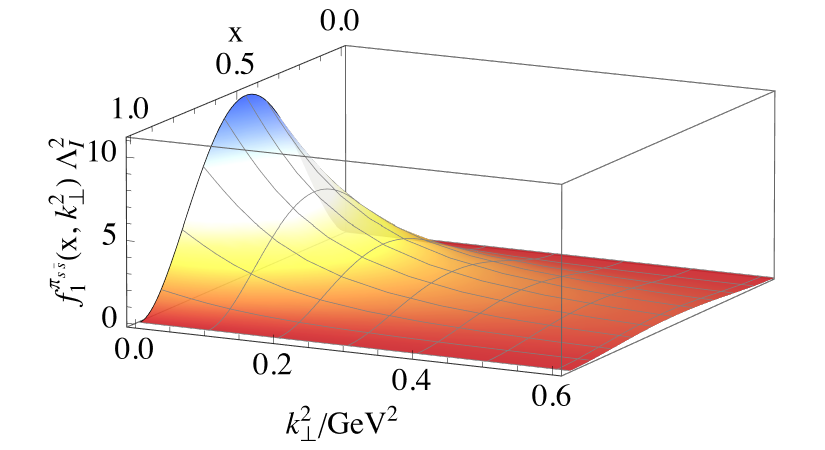}
\end{tabular}
\caption{\label{Figf13D}
Helicity-independent TMDs obtained from LFWFs in Fig.\,\ref{fig:LFWF}.
In each case, normalisation guarantees Eq.\,\eqref{EqBNcons}.
{\sf Panel A}.  $\pi$ RL.
{\sf Panel B}.  $\pi$ bRL.
{\sf Panel C}.  $\pi_{s \bar s}$ RL.
{\sf Panel D}.   $\pi_{s \bar s}$ bRL.
}
\end{figure*}

To understand and resolve the ambiguity highlighted in (\textit{iii}), it is necessary to consider the $k_\perp^2$ dependence on a larger domain; see Fig.\,\ref{F4T}.
Considering these profiles, obtained after rescaling such that $\psi_{\mathsf n}^0(1/2,0)$ is unity for each quark current mass, one then observes the following.
\begin{enumerate}[label=(\textit{\alph*})]
\item Compare the $k_\perp^2$-profiles of $\psi_{\mathsf n}^0(x,k_\perp^2)$.
Plainly, RL truncation produces a more dilated function than bRL on the entire $k_\perp^2$ domain.
Considering \linebreak $|k_\perp| \psi_{\mathsf n}^1(x,k_\perp^2)$, on the other hand, RL truncation yields narrower (contracted) $k_\perp^2$ profiles.
Evidently, additional dynamical EHM effects captured by the bRL kernel work to enhance the $k_\perp^2$ support of the $\mathpzc L=1$ component of the LFWF.

\item Turning to the quark current mass dependence,
RL truncation results show large sensitivity to $\hat m$:
for both LFWF components, increasing $\hat m$ leads to a more dilated $k_\perp^2$ profile.
In contrast, bRL predicts a modest dilation of the $\hat m_s$ $\psi_{\mathsf n}^0(x,k_\perp^2)$ profile when compared with that of the kindred $\hat m$ amplitude: evidently, with the enhancement of EHM, the relative impact of Higgs-generated mass is smaller.

More strikingly, bRL delivers an inversion of the $|k_\perp| \psi_{\mathsf n}^1(x,k_\perp^2)$ profile ordering when compared with the RL results.
This further highlights the impact of EHM in the light quark sector: bRL enhances light-quark orbital angular momentum on $k_\perp^2 \gtrsim (0.6\,{\rm GeV})^2$.


\end{enumerate}
Given Eq.\,\eqref{DefTMD}, then one should expect EHM, expressed via the bRL kernel, to have an impact on the $k_\perp^2$ profile of the TMD, with the importance of the $\mathpzc L=1$ contribution being significantly amplified.

\section{TMDs}
\label{Sec5}
Using the LFWFs in Fig.\,\ref{fig:LFWF} and Eq.\,\eqref{DefTMD}, one obtains the TMDs depicted in Fig.\,\ref{Figf13D}.
Remarks (\textit{i})\,--\,(\textit{iii}) and (\textit{a}), (\textit{b}) in Sec.\,\ref{Sec4} are also qualitatively applicable to the TMDs.

Analyses in perturbative QCD predict \cite{Lepage:1980fj} that, on $ k_\perp^2 \gg m_N^2$, $\psi_{\mathsf 5}^{0}(x,k_\perp^2)$ and $k_\perp^2 \psi_{\mathsf 5}^{1}(x,k_\perp^2)$ behave as $1/k_\perp^2$ up to (damping) $\ln k_\perp^2$ corrections.
Such behaviour is also expressed in the CSM results described above; hence, the meson TMDs behave as follows:
\begin{equation}
f_1(x,k_\perp^2) \stackrel{k_\perp^2 \gg m_N^2}{\propto}
\frac{1}{k_\perp^4 (\ln k_\perp^2)^\gamma}\,,
\end{equation}
where $0<\gamma<1$ is an anomalous dimension.  Whilst
$\int dx \int d^2k_\perp f_1(x,k_\perp^2) = 1 <\infty $ in such circumstances, the mean $k_\perp^2$ integral is not, \textit{viz}.
\begin{equation}
    \langle k_\perp^2 \rangle
    = \int dx \, \int d^2 k_\perp \, k_\perp^2 f_1(x,k_\perp^2)
    \stackrel{\rm QCD}{=} \infty\,.
\end{equation}
Notwithstanding that, if one were to find that an exponential in $k_\perp^2$ (often called Gaussian) did provide a good approximation to the TMD on $k_\perp^2 \lesssim m_N^2$ -- as is common in phenomenology \cite{Boussarie:2023izj}, then one could define an effective mean-$k_\perp^2$ as follows,
with ${\mathsf N}_{0} = \int dx f_1(x,k_\perp^2=0) $:
\begin{equation}
\langle k_\perp^2 \rangle_{\rm eff}
=
\left.\left[ - \frac{1}{{\mathsf N}_{0}}
\int dx \frac{d}{dk_\perp^2} f_1(x,k_\perp^2)  \right]^{-1} \right|_{k_\perp^2=0} ,
\label{EffectiveMean}
\end{equation}
irrespective of the true large-$k_\perp^2$ behaviour.

Returning to the TMDs in Fig.\,\ref{Figf13D}, it is worth identifying the domain on which a $k_\perp^2$-exponential dependence may serve as a valid approximation.  To that end, we have identified the values of $k_\perp^2$ at which the relative difference between a low-$k_\perp^2$ exponential fit is accurate in ratio to within 20\% and to within a factor of $2$.  The domains are bounded above by the values listed in the first two rows of the following array:
\begin{equation}
\label{kperp2table}
\begin{array}{l|llll}
     &  \pi^{\rm RL} & \pi^{\rm bRL}  &  \pi_{s \bar s}^{\rm RL} & \pi_{s\bar s}^{\rm bRL}  \\\hline
\surd \langle k_\perp^2 \rangle_{20\%}/{\rm GeV} & 0.89 & 0.58 & 1.23 & 0.59 \\
\surd \langle k_\perp^2 \rangle_{100\%}/{\rm GeV}& 1.09 & 0.75 & 1.44 & 0.77 \\\hline
(4/\pi)\langle |k_\perp|\rangle^2/\langle k_\perp^2\rangle
& 0.98 & 0.93 & 0.99 & 0.93 \\\hline
\surd \langle k_\perp^2 \rangle_{\rm eff}^{\mathpzc L=0}/{\rm GeV} & 0.36 & 0.28 & 0.47 & 0.31 \\
\surd \langle k_\perp^2 \rangle_{\rm eff}/{\rm GeV} & 0.45 & 0.37 & 0.55 & 0.37 \\\hline
\end{array}
\end{equation}
so, whilst RL truncation results support a $k_\perp^2$-exponential representation over an integration domain that is potentially useful for phenomenology, the bRL predictions indicate a far more limited domain of validity.

In phenomenology, a different measure is commonly used to estimate the reliability of an exponential in $k_\perp^2$ \textit{Ansatz}.
Namely, were such a profile to be a sound representation, then one would have $\langle |k_\perp|\rangle^2 = (\pi/4) \langle k_\perp^2\rangle$.
This measure is explicated in the third row of Eq.\,\eqref{kperp2table}.
(The results were estimated by integrating on $0\leq k_\perp^2 \leq k_{\perp\,{\rm mf}}^2$.)
Evidently, consistent with the first two rows, whilst RL truncation supports such an \textit{Ansatz} over a domain of relevance to phenomenology, bRL does not.
Given that bRL is the more realistic truncation, then an exponential in $k_\perp^2$ TMD is not supported by direct calculations.

The bottom two rows of the array in Eq.\,\eqref{kperp2table} record values of the effective mean-$k_\perp^2$, defined in Eq.\,\eqref{EffectiveMean}.
Row~4 lists the results obtained using only the ${\mathpzc L}=0$ component of the LFWFs and row~5 is the complete result, calculated with both angular momentum pieces.
The impact of the ${\mathpzc L}=1$ component is larger for light quarks and largest (32\% increase) for the $\pi$ in bRL truncation.

%
Regarding RL results, the inflation of the $k_\perp^2$ domain of support seen in Fig.\,\ref{Figf13D}\,C relative to that in Fig.\,\ref{Figf13D}\,A is expressed in a larger mean $k_\perp^2$.
Equally, the similar support ranges of the bRL $\pi$ and $\pi_{s \bar s}$ TMDs translate into practically identical mean-$k_\perp^2$ values.

It is worth looking closer at the $k_\perp^2$ dependence.
Considering also momenta $k_\perp^2 > 0.6\, {\rm GeV}^2$, \textit{viz}.\ the domain $0<k_\perp^2\leq k_{\perp\,{\rm mf}}^2$, with $k_{\perp\,{\rm mf}}^2$ being the maximum momentum-squared used in fitting Eq.\,\eqref{RealSep} -- see \ref{AppFit}, one finds that the ratio \linebreak
$f_1^\pi(x,k_\perp^2)/f_1^{\pi_{s\bar s}}(x,k_\perp^2)$
decreases uniformly with increasing $k_\perp^2$ when RL LFWFs are used.
%
%
In contrast, with bRL LFWFs on this domain, $f_1^\pi(x,k_\perp^2)/f_1^{\pi_{s\bar s}}(x,k_\perp^2)$
increases uniformly, with increasing $k_\perp^2$, at first slowly, consistent with the mean-$k_\perp^2$ values in Eq.\,\eqref{kperp2table}.
%
Thus, as signalled by the discussion of the LFWFs in Sec.\,\ref{Sec4}, RL and bRL predict different relative behaviour between $\pi$ and $\pi_{s\bar s}$ TMDs on the available domain of precise $k_\perp^2$ results:
RL sees EHM strength diluted by Higgs-boson mass effects even at slightly inflated light-quark masses;
whereas the bRL kernel predicts that EHM predominates over Higgs-boson effects on a much larger domain of current masses.
This is largely because EHM expands the $k_\perp^2$ domain on which the ${\mathpzc L}=1$ component is dominant,
increases its magnitude,
and inverts the ordering of that magnitude in $\pi$ \textit{cf}.\ $\pi_{s\bar s}$.

\section{Summary and Perspective}
\label{Sec6}
Working from Poincar\'e-covariant Bethe-Salpeter wave functions for the pion and a fictitious partner state, $\pi_{s\bar s}$, in which the valence degrees of freedom both have current masses around $0.1\,$GeV, we delivered predictions for the light-front wave functions (LFWFs) of these systems.
In each case, the Bethe-Salpeter wave function was projected onto the light front and the pointwise behaviour of the LFWF was reconstructed from its six lowest-order $k_\perp^2$-dependent Mellin moments,
$\langle x^m\rangle(k_\perp^2)$, $m=0,1,\ldots,5$.
Both components, valence spin-antialigned ($\mathpzc L=0$) and valence spin-aligned ($\mathpzc L=1$), were computed and combined to produce the complete LFWF for each bound state.

In completing this analysis, two distinct Bethe-Sal\-peter kernels were used.
Namely, that obtained at leading order in a widely used symmetry-preserving, systematic truncation scheme (rainbow-ladder, RL) and a novel form (bRL), constructed nonperturbatively, which incorporates dynamical effects driven by physics underlying the emergence of hadron mass (EHM).
By many measures, the bRL kernel provides the more realistic representation of meson structure.

The following features of the LFWFs are worth highlighting.
%
Considering the ratios
$\langle x^m\rangle(k_\perp^2)/\langle x^0 \rangle(k_\perp^2)$, $m=1,\,\ldots,5$, for both LFWF spin projections [Fig.\,\ref{FmomentsSep}], then when calculated using the RL kernel, they are all approximately $k_\perp^2$-independent, with each taking a value that is a fair match with the associated moment of the system's distribution amplitude (DA).
Using instead the bRL kernel, the ratios are constants which are practically indistinguishable from the related DA moments.
These outcomes are true for both systems.
Hence, their LFWFs can be written in a separable form [Eq.\,\eqref{EqSep}, \eqref{RealSep}], with this being a reasonable approximation for RL results and a pointwise reliable representation of bRL LFWFs.

Regarding the reconstructed LFWFs themselves \linebreak 
$\,$ [Fig.\,\ref{fig:LFWF}], the $\mathpzc L=1$ component is important in all cases.
Therefore, its omission delivers a poor approximation.
Significant misrepresentations can nevertheless be avoid\-ed in integrated quantities, like collinear distribution functions, because both LFWF spin projections are separable, with the dependence on light-front $x$ being the same in both cases.
Notably, the realistic bRL kernel produces much stronger $\mathpzc L=1$ components in both systems [Fig.\,\ref{F4T} and associated discussion].

Naturally, these features of the LFWF are expressed in the associated helicity-independent transverse momentum dependent distribution functions (TMDs) \linebreak $\,$ [Sec.\,\ref{Sec5}].
Of particular note are the impacts of the bRL kernel.
Whereas RL results are compatible with an exponential in $k_\perp^2$ (Gaussian) \textit{Ansatz} because EHM effects are underestimated, the realistic bRL kernel delivers a more complex LFWF to which a Gaussian provides, at best, only a rough guide to the pointwise $k_\perp^2$ dependence.
The phenomenological reliability  of Gaussian \textit{Ans\"atze} should therefore not be too greatly depended upon.

With a well-grounded pion LFWF in hand, one may now, \textit{e.g}, revisit existing, exploratory studies of the pion Boer-Mulders (BM) function \cite{Boer:1997nt} and work toward a realistic picture of the dependence of pion observables on the transverse spin of its valence constituents.
Along the way, one might also learn about the true efficacy of eikonal approximations in estimating the magnitude of the final-state interactions between struck and spectator degrees-of-freedom that are necessary to obtain a nonzero BM function.
Such information could be valuable in understanding analogous features of, and influences on, baryon structure.

\begin{acknowledgements}
We thank K.\ Raya for constructive comments.
Work supported by:
National Natural Science Foundation of China, grant no.\ 12135007;
Helmholtz-Zentrum Dres\-den-Rossendorf, under the High Potential
Programme;
and
Ministerio Espa\~nol de Ciencia, Innovaci\'on y Universidades (MICINN) grant no.\ PID2022-140440NB-C22.

\medskip

\noindent\textbf{Data Availability Statement} Data will be made available on reasonable request.  [Authors' comment: All information necessary to reproduce the results described herein is contained in the material presented above.]
\medskip

\noindent\textbf{Code Availability Statement} Code/software will be made available
on reasonable request. [Authors' comment: No additional remarks.]

\end{acknowledgements}

\appendix
\section{Fitting procedure and uncertainty analysis}
\label{AppFit}
For each system considered, one has a set of $k_\perp^2$-depen\-dent Mellin moment curves, $\{(j,\langle x^{j}\rangle(k_\perp^2))\}_{j=0}^{N_x}$.
We use $N_x=5$.
In fitting, we introduce a discrete set of $N_k=40$ $k_\perp^2$ values, evenly spaced on $0< k_\perp^2/{\rm GeV}^2\leq 3=: k_{\perp\,{\rm mf}}^2$.  (We have established that $N_k=40$ is sufficient to obtain converged results.)
We then choose to minimise the following $\chi^2$ function:
{\allowdisplaybreaks
\begin{subequations}
\label{eq:chi2}
\begin{align}
    &\chi^2(\vec{\theta}
  =\{a_0^{\mathpzc L},b_1^{\mathpzc L}, b_2^{\mathpzc L}, c_0^{\mathpzc L}, d_1^{\mathpzc L}, {\mathpzc m}_2^{\mathpzc L}, {\mathpzc m}_3^{\mathpzc L}, {\mathpzc m}_4^{\mathpzc L}, {\mathpzc m}_5^{\mathpzc L}\}) \nonumber \\
  &= \sum_{i}^{N_k} \sum_{j=2}^{N_x}
    \left[
    \int_0^1 dx\, x^j \psi_{\mathsf 5}^{\mathpzc L}(x,k^2) - \langle x^j\rangle(k^2)\right]^2\\
    & \stackrel{\rm separable}{=}
    \sum_{i}^{N_k} \sum_{j=2}^{N_x}
   \left[ {\mathpzc p}_{\mathsf 5}^{\mathpzc L}(k^2) \int_0^1 dx\, x^j\varphi_{\mathsf 5}^{\mathpzc L}(x)
   - \langle x^j\rangle^{\mathpzc L}(k^2)\right]^2\\
    & = \sum_{i}^{N_k} \sum_{j=2}^{N_x}
     \left[{\mathpzc p}_{\mathsf 5}^{\mathpzc L}(k^2) {\mathpzc m}_j^{\mathpzc L} -
     \langle x^0\rangle^{\mathpzc L}(k^2) \frac{\langle x^j \rangle^{\mathpzc L}(k^2)}{\langle x^0\rangle^{\mathpzc L}(k^2)}\right]^2, \label{EqA1c}
\end{align}
\end{subequations}
where we have introduced
\begin{equation}
{\mathpzc m}_j^{\mathpzc L} = \int_0^1 dx \, x^j\, \varphi_{\mathsf 5}^{\mathpzc L}(x)\,
\label{Moments}
\end{equation}
and the sum begins at $j=2$ because $\mathpzc m_0=1$ (DA normalisation) and  $\mathpzc m_1=1/2$ (DA symmetry under $x\leftrightarrow [1-x])$.
For the pion, the moment ratios in Eq.\,\eqref{EqA1c} are drawn in Fig.\,\ref{FmomentsSep}.}

\textit{N.B}.\ Although Fig.\,\ref{FmomentsSep} shows that, for all intents and purposes, the LFWF $x$-profiles satisfy $\varphi_{\mathsf 5}^{0}(x) = \varphi_{\mathsf 5}^{1}(x)$, in this Appendix, we elect to provide additional numerical validation of that result by treating the moments in Eq.\,\eqref{Moments} as ${\mathpzc L}$-dependent fitting parameters.  As we shall find -- see Table~\ref{tab:momentspi} below, the returned best-fit values satisfy ${\mathpzc m}_j^0={\mathpzc m}_j^1$ within mutual uncertainties.  With this, one confirms $\varphi_{\mathsf 5}^{0}(x) = \varphi_{\mathsf 5}^{1}(x)$.

Hereafter, one has a sum of many squares, which must be minimised.
In this $\chi^2$ minimisation process, there are nine parameters for ${\mathpzc L=0}$ and seven for ${\mathpzc L=1}$.
We denote the results by
\begin{equation}
\vec{\theta} = {\vec\theta}_{\rm f}=
(a_{0f}^{\mathpzc L},b_{1f}^{\mathpzc L}, b_{2f}^{\mathpzc L}, c_{0f}^{\mathpzc L}, d_{1f}^{\mathpzc L}, {\mathpzc m}_{2f}^{\mathpzc L}, {\mathpzc m}_{3f}^{\mathpzc L}, {\mathpzc m}_{4f}^{\mathpzc L}, {\mathpzc m}_{5f}^{\mathpzc L})
\end{equation}
and list their values in Tables~\ref{tab:paramspi}, \ref{tab:momentspi}.
Working with the values in Table~\ref{tab:momentspi}, one obtains the results for $\rho_{\mathsf 5}^{\mathpzc L}$ listed in Table~\ref{tab:paramspi} by simple one-variable minimisation of moment comparisons.

The Hessian matrix associated with the above $\chi^2$ minimisation procedure is
\begin{equation}
\left[ {\mathpzc H}_{ij} \right] =
\left.
\left[ \frac{\partial}{\partial \theta_i}\frac{\partial}{\partial \theta_j}
\chi^2(\theta) \right]\right|_{\theta=\theta_{\rm f}}\,.
\end{equation}
As usual, with the Hessian in hand, we take the uncertainty in each fitting parameter to be
\begin{equation}
    \sigma_i = \sqrt{[{\mathpzc H}^{-1}]_{ii}}\,,
    \quad {\rm no~sum~on~}i\,,
\end{equation}
and the associated correlation matrix is obtained as
\begin{equation}
    [{\mathpzc C}]_{ij} = [H^{-1}]_{ij}/\sqrt{[H^{-1}]_{ii}[H^{-1}]_{jj}}\,,
\end{equation}
again with no sum on repeated indices.

\begin{table}[t]
\centering
\caption{%
Moments obtained via minimisation of Eq.\,\eqref{eq:chi2} for the  truncations and systems considered herein.
The last row lists these moments for the asymptotic DA.
Associated correlation matrices are listed in Eqs.\,\eqref{CMI}\,--\,\eqref{CMF}.
\label{tab:momentspi}}
\begin{tabular*}
{\hsize}
{
l@{\extracolsep{0ptplus1fil}}
|l@{\extracolsep{0ptplus1fil}}
l@{\extracolsep{0ptplus1fil}}
l@{\extracolsep{0ptplus1fil}}
l@{\extracolsep{0ptplus1fil}}}\hline\hline
\centering
$\varphi_{{\mathsf 5}R}^{\mathpzc L}\ $ & ${\mathpzc m}_2^{{\mathpzc L}}$ & ${\mathpzc m}_3^{{\mathpzc L}}$ & ${\mathpzc m}_4^{{\mathpzc L}}$& ${\mathpzc m}_5^{{\mathpzc L}}$ \\
\hline
${\psi}^{\uparrow\downarrow}_{\pi,\,\mathrm{RL}}\ $&
$0.3171(42)$ & $0.2258(41)$&$0.1721(41)$&$0.1372(41)$ \\
${\psi}^{\uparrow\uparrow}_{\pi,\,\mathrm{RL}}\ $&
$0.3217(24)$ & $0.2325(24)$&$0.1809(24)$&$0.1481(24)$\\
${\psi}^{\uparrow\downarrow}_{\pi,\,\mathrm{bRL}}\ $&
$0.3096(31)$ & $0.2144(31)$&$0.1594(31)$&$0.1244(30)$ \\
${\psi}^{\uparrow\uparrow}_{\pi,\,\mathrm{bRL}}\ $&
$0.3077(13)$&$0.2116(13)$&$0.1561(13)$&$0.1208(13)$ \\
\hline
${\psi}^{\uparrow\downarrow}_{\pi_{s \bar s},\,\mathrm{RL}}\ $&$0.3008(49)$&$0.2011(48)$&$0.1441(47)$&$0.1084(47)$\\
${\psi}^{\uparrow\uparrow}_{{\pi_{s \bar s},\,\mathrm{RL}}}\ $&
$0.2997(42)$&$0.1995(41)$&$0.1421(41)$&$0.1061(41)$\\
${\psi}^{\uparrow\downarrow}_{{\pi_{s \bar s},\,\mathrm{bRL}}}\ $
&$0.2983(32)$&$0.1976(31)$&$0.1403(31)$&$0.1047(31)$ \\
${\psi}^{\uparrow\uparrow}_{{\pi_{s \bar s},\,\mathrm{bRL}}}\ $
&$0.2964(17)$&$0.1946(17)$&$0.1370(17)$&$0.1013(17)$ \\
\hline\hline
$\varphi_{\rm as}$ &
$0.3$ &  $0.2$ & $0.143$ & $0.107$\\\hline\hline
\end{tabular*}
\end{table}

The uncertainties on each reconstruction parameter are listed in Tables~\ref{tab:paramspi}, \ref{tab:momentspi}.
Regarding Table~\ref{tab:momentspi}, within any reasonable understanding of the precision of our results, there is no meaningful difference between $\pi_{s\bar s}$ moments and those of $\varphi_{\rm as}$.
This outcome confirms the analysis in Refs.\,\cite{Ding:2015rkn, Chen:2018rwz, Ding:2018xwy}.

Calculated results for the correlation matrices follow.  In all cases, the result is block diagonal:
\begin{equation}
\label{eq:Rmatpirl0}
\begin{array}{c|cc}
{\psi}^{\mathpzc L}_{\mathsf 5} & &
\\\hline
& {\mathpzc C}^{\mathpzc L}_{{{\mathsf 5}{\mathsf p}}}  & 0 \\
& 0 & {\mathpzc C}^{\mathpzc L}_{{{\mathsf 5}{\mathsf m}}} \\
\end{array}
\end{equation}
One will see that the moments in Eq.\,\eqref{Moments} are practically linearly independent.

\subsection{Pion}

\begin{equation}
\label{CMI}
\begin{array}{c|rrrrr}
{\mathpzc C}_{\pi\mathpzc p, {\rm RL}}^{\uparrow\downarrow}
& a_{0}^{0} & b_{1}^{0} & b_{2}^{0} & c_{0}^{0} & d_{1}^{0}\\\hline
a_{0}^{0} & 1 & -0.92  & -0.50 &  -0.99 & 0.81  \\
b_{1}^{0} & -0.92 & 1  & 0.72 & 0.92     & -0.60  \\
b_{2}^{0} & -0.50 & -0.72 & 1  & 0.50     & -0.26  \\
c_{0}^{0} & -0.99 & 0.92 & 0.50  & 1 & -0.78   \\
d_{1}^{0} & 0.81 & 0.60  & -0.26 & -0.78    & 1  \\
\end{array}
\end{equation}

\begin{equation}
\begin{array}{c|rrrr}
{\mathpzc C}_{\pi\mathpzc m, {\rm RL}}^{\uparrow\downarrow} &
{\mathpzc m}_2^0 & {\mathpzc m}_3^0 & {\mathpzc m}_4^0 & {\mathpzc m}_5^0 \\\hline
{\mathpzc m}_2^0 & 1   & 0.05  & 0.04 & 0.03 \\
{\mathpzc m}_3^0 & 0.05   & 1   & 0.03  & 0.02  \\
{\mathpzc m}_4^0 & 0.04 & 0.03  & 1   & 0.02    \\
{\mathpzc m}_5^0  & 0.03   & 0.02      & 0.02 & 1 \\
\end{array}
\end{equation}

\begin{equation}
\begin{array}{c|rrr}
{\mathpzc C}_{\pi\mathpzc p, {\rm RL}}^{\uparrow\uparrow}
& a_{0}^{1} & b_{1}^{1} & b_{2}^{1} \\\hline
a_{0}^{1} & 1 & 0.54  & -0.22    \\
b_{1}^{1} & 0.54 & 1  & -0.76   \\
b_{2}^{1} & -0.22 & -0.76 & 1   \\
\end{array}
\end{equation}

\begin{equation}
\begin{array}{c|rrrr}
{\mathpzc C}_{\pi\mathpzc m, {\rm RL}}^{\uparrow\uparrow}
&
{\mathpzc m}_2^1 & {\mathpzc m}_3^1 & {\mathpzc m}_4^1 & {\mathpzc m}_5^1 \\\hline
{\mathpzc m}_2^1 & 1   & 0.06  & 0.04 & 0.04 \\
{\mathpzc m}_3^1 & 0.06   & 1   & 0.03  & 0.03  \\
{\mathpzc m}_4^1 & 0.04 & 0.03  & 1   & 0.02    \\
{\mathpzc m}_5^1  & 0.04   & 0.03      & 0.02 & 1 \\
\end{array}
\end{equation}

\begin{equation}
\begin{array}{c|rrrrr}
{\mathpzc C}_{\pi\mathpzc p, {\rm bRL}}^{\uparrow\downarrow}
& a_{0}^{0} & b_{1}^{0} & b_{2}^{0} & c_{0}^{0} & d_{1}^{0}\\\hline
a_{0}^{0} & 1 & 0.92  & -0.50 &  -0.99 & 0.82  \\
b_{1}^{0} & -0.92 & 1  & -0.70 & -0.92     & 0.62  \\
b_{2}^{0} & -0.50 & -0.70 & 1  & 0.50     & -0.27  \\
c_{0}^{0} & -0.99 & 0.92 & 0.50  & 1 & -0.79   \\
d_{1}^{0} & 0.82 & 0.62  & -0.27 & -0.79    & 1  \\
\end{array}
\end{equation}

\begin{equation}
\begin{array}{c|rrrr}
{\mathpzc C}_{\pi\mathpzc m, {\rm bRL}}^{\uparrow\downarrow}
&
{\mathpzc m}_2^0 & {\mathpzc m}_3^0 & {\mathpzc m}_4^0 & {\mathpzc m}_5^0 \\\hline
{\mathpzc m}_2^0 & 1   & 0.05  & 0.04 & 0.03 \\
{\mathpzc m}_3^0 & 0.05   & 1   & 0.03  & 0.02  \\
{\mathpzc m}_4^0 & 0.04 & 0.03  & 1   & 0.01    \\
{\mathpzc m}_5^0  & 0.03   & 0.02      & 0.01 & 1 \\
\end{array}
\end{equation}

\begin{equation}
\begin{array}{c|rrr}
{\mathpzc C}^{\uparrow\uparrow}_{\pi \mathpzc p \mathrm{bRL}}
& a_{0}^{1} & b_{1}^{1} & b_{2}^{1} \\\hline
a_{0}^{1} & 1 & 0.54  & -0.21    \\
b_{1}^{1} & 0.54 & 1  & -0.72   \\
b_{2}^{1} & -0.21 & -0.72 & 1   \\
\end{array}
\end{equation}

\begin{equation}
\begin{array}{c|rrrr}
{\mathpzc C}^{\uparrow\uparrow}_{\pi \mathpzc m \mathrm{bRL}}
&
{\mathpzc m}_2^1 & {\mathpzc m}_3^1 & {\mathpzc m}_4^1 & {\mathpzc m}_5^1 \\\hline
{\mathpzc m}_2^1 & 1   & 0.05  & 0.04 & 0.03 \\
{\mathpzc m}_3^1 & 0.06   & 1   & 0.03  & 0.02  \\
{\mathpzc m}_4^1 & 0.04 & 0.03  & 1   & 0.01    \\
{\mathpzc m}_5^1  & 0.03   & 0.02      & 0.01 & 1 \\
\end{array}
\end{equation}


\subsection{Heavy pion}

\begin{equation}
\begin{array}{c|rrrrr}
{\mathpzc C}_{\pi_{s\bar s}\mathpzc p, {\rm RL}}^{\uparrow\downarrow}
& a_{0}^{0} & b_{1}^{0} & b_{2}^{0} & c_{0}^{0} & d_{1}^{0}\\\hline
a_{0}^{0} & 1 & 0.92  & -0.48 &  0.99 & 0.81  \\
b_{1}^{0} & 0.92 & 1  & -0.71 & 0.92     & 0.60  \\
b_{2}^{0} & -0.48 & -0.71 & 1  & -0.48     & -0.25  \\
c_{0}^{0} & -0.99 & 0.92 & -0.48  & 1 & 0.77   \\
d_{1}^{0} & 0.81 & 0.60  & -0.25 & 0.77    & 1  \\
\end{array}
\end{equation}

\begin{equation}
\begin{array}{c|rrrr}
{\mathpzc C}_{\pi_{s\bar s}\mathpzc m, {\rm RL}}^{\uparrow\downarrow}
&
{\mathpzc m}_2^0 & {\mathpzc m}_3^0 & {\mathpzc m}_4^0 & {\mathpzc m}_5^0 \\\hline
{\mathpzc m}_2^0 & 1   & 0.05  & 0.03 & 0.02 \\
{\mathpzc m}_3^0 & 0.05   & 1   & 0.02  & 0.02  \\
{\mathpzc m}_4^0 & 0.03 & 0.02  & 1   & 0.01    \\
{\mathpzc m}_5^0  & 0.02   & 0.02      & 0.01 & 1 \\
\end{array}
\end{equation}

\begin{equation}
\begin{array}{c|rrr}
{\mathpzc C}_{\pi_{s\bar s}\mathpzc p, {\rm RL}}^{\uparrow\uparrow}
& a_{0}^{1} & b_{1}^{1} & b_{2}^{1} \\\hline
a_{0}^{1} & 1 & 0.55  & -0.21    \\
b_{1}^{1} & 0.55 & 1  & -0.75   \\
b_{2}^{1} & -0.21 & -0.75 & 1   \\
\end{array}
\end{equation}

\begin{equation}
\begin{array}{c|rrrr}
{\mathpzc C}_{\pi_{s\bar s}\mathpzc m, {\rm RL}}^{\uparrow\uparrow}
&
{\mathpzc m}_2^1 & {\mathpzc m}_3^1 & {\mathpzc m}_4^1 & {\mathpzc m}_5^1 \\\hline
{\mathpzc m}_2^1 & 1   & 0.05  & 0.03 & 0.02 \\
{\mathpzc m}_3^1 & 0.05   & 1   & 0.02  & 0.02  \\
{\mathpzc m}_4^1 & 0.03 & 0.02  & 1   & 0.01    \\
{\mathpzc m}_5^1  & 0.02   & 0.02      & 0.01 & 1 \\
\end{array}
\end{equation}

\begin{equation}
\begin{array}{c|rrrrr}
{\mathpzc C}_{\pi_{s\bar s}\mathpzc p, {\rm bRL}}^{\uparrow\downarrow}
& a_{0}^{0} & b_{1}^{0} & b_{2}^{0} & c_{0}^{0} & d_{1}^{0}\\\hline
a_{0}^{0} & 1 & 0.92  & 0.50 &  -0.99 & 0.82  \\
b_{1}^{0} & 0.92 & 1  & 0.71 & -0.92     & 0.62  \\
b_{2}^{0} & 0.51 & 0.71 & 1  & -0.51     & 0.27  \\
c_{0}^{0} & -0.99 & -0.92 & -0.51  & 1 & -0.79   \\
d_{1}^{0} & 0.82 & 0.61  & 0.27 & -0.79    & 1  \\
\end{array}
\end{equation}

\begin{equation}
\begin{array}{c|rrrr}
{\mathpzc C}_{\pi_{s\bar s}\mathpzc m, {\rm bRL}}^{\uparrow\downarrow}
&
{\mathpzc m}_2^0 & {\mathpzc m}_3^0 & {\mathpzc m}_4^0 & {\mathpzc m}_5^0 \\\hline
{\mathpzc m}_2^0 & 1   & 0.04  & 0.03 & 0.02 \\
{\mathpzc m}_3^0 & 0.04   & 1   & 0.02  & 0.02  \\
{\mathpzc m}_4^0 & 0.03 & 0.02  & 1   & 0.01    \\
{\mathpzc m}_5^0  & 0.02   & 0.02      & 0.01 & 1 \\
\end{array}
\end{equation}

\begin{equation}
\begin{array}{c|rrr}
{\mathpzc C}_{\pi_{s\bar s}\mathpzc p, {\rm bRL}}^{\uparrow\uparrow}
& a_{0}^{1} & b_{1}^{1} & b_{2}^{1} \\\hline
a_{0}^{1} & 1 & 0.55  & -0.22    \\
b_{1}^{1} & 0.55 & 1  & -0.74   \\
b_{2}^{1} & -0.22 & -0.74 & 1   \\
\end{array}
\end{equation}

\begin{equation}
\begin{array}{c|rrrr}
\label{CMF}
{\mathpzc C}_{\pi_{s\bar s}\mathpzc m, {\rm bRL}}^{\uparrow\uparrow}
&
{\mathpzc m}_2^1 & {\mathpzc m}_3^1 & {\mathpzc m}_4^1 & {\mathpzc m}_5^1 \\\hline
{\mathpzc m}_2^1 & 1   & 0.04  & 0.03 & 0.02 \\
{\mathpzc m}_3^1 & 0.04   & 1   & 0.02  & 0.02  \\
{\mathpzc m}_4^1 & 0.03 & 0.02  & 1   & 0.01    \\
{\mathpzc m}_5^1  & 0.02   & 0.02      & 0.01 & 1 \\
\end{array}
\end{equation}


\end{document}